\documentclass[twocolumn,trackchanges]{aastex631}
\turnoffeditone 
\usepackage{amsmath,natbib}
\received{November 17, 2022}
\accepted{January 31, 2023}

\submitjournal{ApJ}

\newcommand{\cii}{[\ion{C}{2}]}
\newcommand{\oiii}{[\ion{O}{3}]}
\newcommand{\niii}{[\ion{N}{3}]}
\newcommand{\oi}{[\ion{O}{1}]}
\makeatletter
\newcommand\hii{H\,{%
\ifx\@currsize\normalsize\small \else
\ifx\@currsize\small\footnotesize \else
\ifx\@currsize\footnotesize\scriptsize \else
\ifx\@currsize\scriptsize\tiny \else
\ifx\@currsize\large\normalsize \else
\ifx\@currsize\Large\large
\fi\fi\fi\fi\fi\fi
\rmfamily\@Roman{2}}\relax}\makeatother
\newcommand{\CO}{CO($14\to13$)}
\newcommand{\um}{\mbox{\,$\mu$m}}
\newcommand{\qcm}{\mathrm{cm}^{-3}}
\newcommand{\GUV}{$G_{\mathrm UV}$}
\newcommand{\nG}{$(n,G_{\mathrm UV})$}
\newcommand{\merg}{$\cdot10^{-3}\mathrm{erg\,cm^{-2}s^{-1}sr^{-1}}$}
\newcommand{\ifir}{$I_{\mathrm FIR}$}
\shorttitle{The PDR in M17-SW}
\shortauthors{Klein et al.}

\begin{document}

\title{The PDR fronts in M17-SW localized with FIFI-LS onboard SOFIA}

%% AUTHORS

%% Include each other on their own "\author" command.
%% \author[xxxx-xxxx-xxxx-xxxx]{Author Name}
%% where the the "x" are the ORCID of the author. If you don't have it leave out the "[]" part of the command.

%% The new \correspondingauthor command is available in V6.2 to identify the
%% corresponding author of the manuscript. 
%% Use \email to set provide email addresses. Each \email will appear on its
%% own line so you can put multiple email address in one \email call. 

\correspondingauthor{Randolf Klein}
\email{rklein@sofia.usra.edu}

\author[0000-0002-7187-9126]{Randolf Klein}
\affiliation{SOFIA/USRA, NASA Ames Research Center, P.O. Box 1, MS 232-12, Moffett Field, CA 94035, USA}

\author{Alexander Reedy}
\affiliation{California Institute of Technology, 1200 E. California Blvd, Pasadena, CA 91125, USA}

\author[0000-0003-2649-3707]{Christian Fischer}
\affiliation{Deutsches SOFIA Institut, University of Stuttgart, Pfaffenwaldring 29, D-70569 Stuttgart, Germany}

\author[0000-0002-4540-6587]{Leslie Looney}
\affiliation{Department of Astronomy, University of Illinois, 1002 West Green Street, Urbana, IL 61801, USA}

\author[0000-0002-5613-1953]{Sebastian Colditz}
\affiliation{Deutsches SOFIA Institut, University of Stuttgart, Pfaffenwaldring 29, D-70569 Stuttgart, Germany}

\author[0000-0002-3698-7076]{Dario Fadda}
\affiliation{SOFIA/USRA, NASA Ames Research Center, P.O. Box 1, MS 232-12, Moffett Field, CA 94035, USA}

\author[0000-0003-0306-0028]{Alexander G. G. M. Tielens}
\affiliation{Leiden Observatory, PO Box 9513, 2300 RA Leiden, The Netherlands}
\affiliation{Department of Astronomy, University of Maryland, MD 20742, USA}

\author[0000-0002-9123-0068]{Willam D. Vacca}
\affiliation{SOFIA/USRA, NASA Ames Research Center, P.O. Box 1, MS 232-12, Moffett Field, CA 94035, USA}

\begin{abstract}

  To understand star formation rates, studying feedback mechanisms that regulate star formation is necessary. The radiation emitted by nascent massive stars play a significant role in feedback by photo-dissociating and ionizing their parental molecular clouds. To gain a detailed picture of the physical processes, we mapped the photo-dissociation region (PDR) M17-SW in several fine structure and high-J CO lines with FIFI-LS, the far-infrared \edit1{imaging} spectrometer aboard SOFIA. An analysis of the CO and [O I]146$\mu$m line intensities, combined with the far infrared intensity, allows us to create a density and UV intensity map using a one dimensional model. The density map reveals a sudden change in the gas density crossing the PDR. The strengths and limits of the model and the locations of the ionization and photo-dissociation front of the edge-on PDR are discussed. 

\end{abstract}

%% and \citet commands to identify citations.  

\section{Introduction}
\label{sec:intro}

Massive stars play a major role in shaping our Galaxy and other galaxies due to their energy output throughout their lifetime. One aspect is their impact on the interstellar medium (ISM) due to their energetic and ionizing radiation. As massive stars form inside dense molecular clouds, parental clouds gets eroded by the stellar radiation. In so-called photo-dissociation  or photon-dominated  regions (PDRs, e.g.\ \citealt{Hollenbach97,Hollenbach99}), the gas becomes ionized, photo-dissociated, and heated via  the photo-electric effect on poly-cyclic aromatic hydrocarbon molecules (PAHs) and very small dust grains. Subsequently, the gas cools through continuum emission from dust grains and line emission, especially atomic and ionic fine structure lines such as \oi, \cii, and \oiii. Understanding the resulting interface between the emerging \hii\ region and the evaporating molecular cloud, as well as their relationship with the PDR is crucial to understand how the ISM is processed and how star formation regulates itself.

An ideal target to study the destruction of molecular clouds by massive stars forming inside them is the well-studied edge-on PDR  M17-SW. We observed several lines  of this object which act as cooling lines or can serve as diagnostics with FIFI-LS. The far-infrared imaging spectrometer FIFI-LS \citep{Fischer2018,Colditz2018} on SOFIA \citep{Temi2018} allowed us to map a large portion of M17-SW in key far infrared transitions with the goal of modeling the whole mapped area with a simple PDR model.

M17 also known in the Simbad database \citep{simbad} as Checkmark, Horseshoe, Lobster, Omega, and  Swan Nebula, is an \hii\ region in the Carina-Sagittarius spiral arm ionized by about 100 OB-stars \citep{Lada1991} in the young ($<10^6$\,yr; \citealt{Hanson}) open cluster NGC\,6618. \cite{Hoffmeister2008} classified spectroscopically 46 OB-stars including 20 O-stars. They also placed M17 at a spectro-photometric distance of $2.1\pm0.2$\,kpc. However, \cite{Kuhn2019} place M17 at a distance of 1.7\,kpc using Gaia DR2 data.

\begin{figure}
  \includegraphics[width=\linewidth]{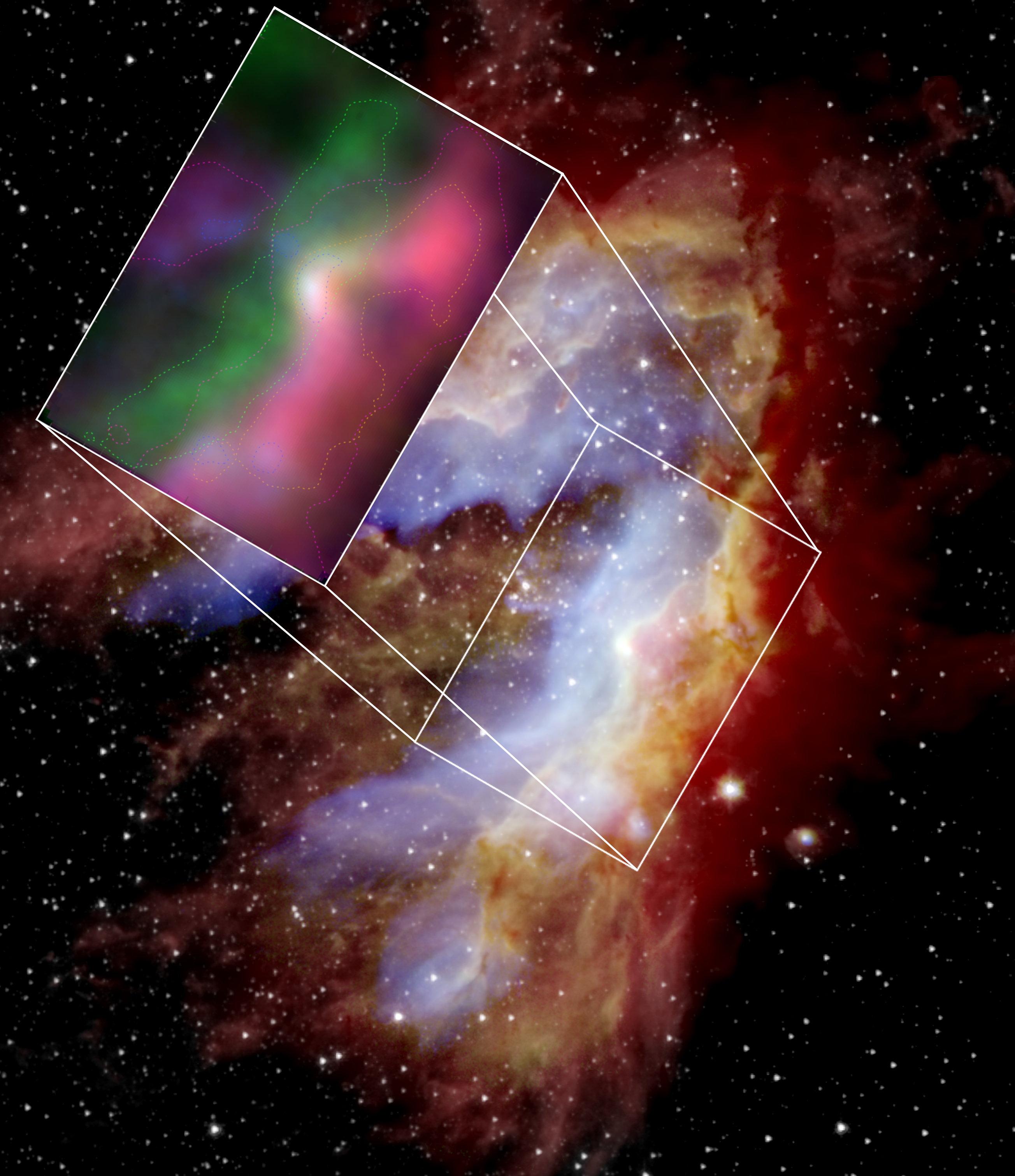}
  \caption{M17-SW Overview: Background - M17 (3.6\um - white, 20\um - blue, 37\um - green, 70\um - red) by \protect\cite{Lim2020}; the box shows the approximate location of the FIFI-LS maps. The inset is also shown in Fig.~\ref{fig:fs_lines}}
  \label{fig:m17-sw}
\end{figure}

The cluster created a large blister \hii\ region open to the south-east. The cavity is filled with hot gas visible in X-rays \citep{Townsley2003}. It splits the giant molecular cloud into two parts, M17-N and M17-SW. The cavity and the two parts of the molecular cloud can be readily identified in the SOFIA/FORCAST \& Herschel/PACS image by \cite{Lim2020} reproduced in Fig.~\ref{fig:m17-sw} providing an overview of M17. PDRs can be found both in M17-N and M17-SW. As many earlier studies \citep[e.g. ][]{Stutzki90,Meixner92}, we focus our study on the more edge-on PDR in M17-SW. Widespread \cii-emission had been analyzed in the above mentioned studies and more recently again by \cite{Perez2012,Perez2015_atomicgas}.

\begin{figure}
  \includegraphics[width=\linewidth,viewport=35 1 435 323]{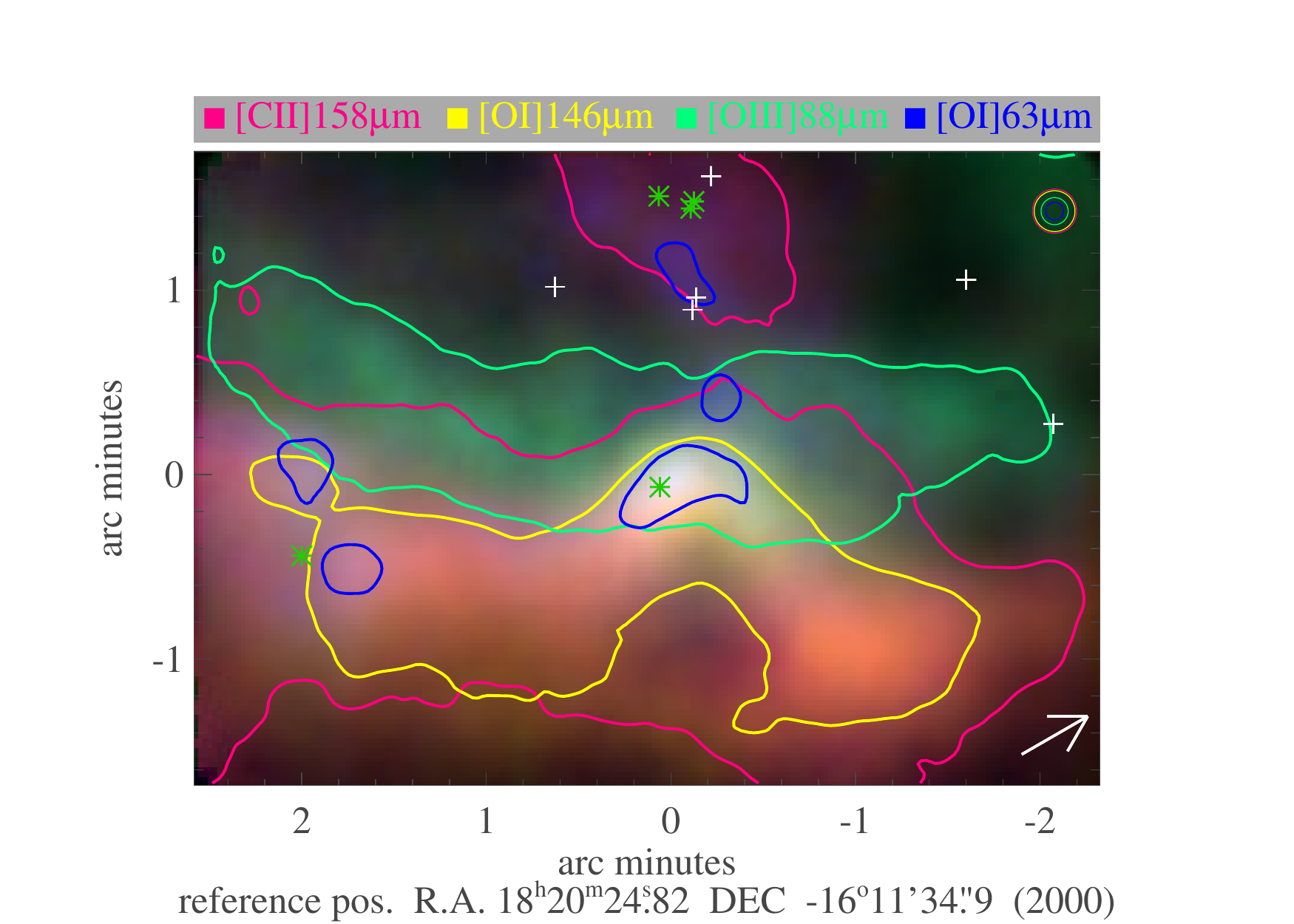}
  \caption{FIFI-LS line intensity maps showing the ionized and PDR layers traced by \oi63\um\ (blue), \oiii88\um\ (green), \oi146\um\ (yellow), and \cii158\um\ (magenta); the colored dashed contours are at 50\% of the peak intensity of the respective line (Tbl.~\ref{tab:values}). The circles in the upper right show the respective beam sizes. The blue symbols indicate the locations of the O-stars identified by \protect\cite{Hoffmeister2008}: stars for spectral types earlier than O9, crosses for types O9 and O9.5.  The arrow points north.}
  \label{fig:fs_lines}
\end{figure}

The box in Fig.~\ref{fig:m17-sw} approximately shows the area mapped by FIFI-LS. The map orientation was chosen, so that the edge-on PDR M17-SW region runs roughly horizontally through the middle of the mapped area \edit1{with the \hii\ region to one side and the molecular cloud to the other side}. Figure~\ref{fig:fs_lines} demonstrates the spatial layering of the observed fine-structure lines due to the edge-on geometry. The reference position for this and for all other maps is the location of the hypercompact \hii\ region M17-UC1 \citep{Sewilo2004}.  The deeply embedded O6\,V-star, M17\,IRS\,5 ($A_V=24.0$, \citealt{Hoffmeister2008}) is located close to M17-UC1 both being embedded in the protrusion extending from the molecular cloud into the \hii\ region indicating on-going star formation in M17-SW. This is also the area where the PDR tracers discussed here peak.

The locations of the O-stars identified by \cite{Hoffmeister2008} in the mapped area are marked in most of our figures, too. The main ionizing sources, CEN1a and CEN1b, two O4\,V-stars are at the top of the mapped area roughly in the middle together with an O6\,V- and an O9\,V-star. O-stars outside of the mapped area are only to the top and the top-right of the mapped area. 

In this paper, we discuss the lines tracing the PDR and their analysis to obtain physical parameters in the PDR. In a forthcoming paper, we will discuss in more detail the lines tracing the \hii\ region and the \hii\ region's physical parameters.

\section{Observations}
\label{sec:observations}

The observations were obtained with FIFI-LS for the open time proposal 04\_0049 (PI: R. Klein). FIFI-LS is an integral-field spectrometer for SOFIA that provides simultaneous observations in two channels: the blue channel covering 51-125 $\mu$m and the red channel covering 115-203 $\mu$m. Each channel consists of an array of $5 \times 5$ pixels covering a field of 30" and 60" for the blue and red channel, respectively. The two fields have a slightly different center (approximately 6\arcsec). The optics within FIFI-LS rearranges the 25 spaxels (spatial pixels) into a pseudo-slit, and the light impinging on each spaxel is then dispersed with a grating along 16 pixels. This generates an integral-field data cube for each observation. The spectral resolution $R = \lambda$/$\Delta\lambda$  is wavelength dependent, ranging from 500 to 2000.

Most observations were obtained on flights from Christchurch, New Zealand, in June and July 2016. Additional data were obtained on a flight from Palmdale, California, in May 2019. The observational details are summarized in Table \ref{tab:obs}. All observations used the D105 dichroic to separate red and blue channels. The FIFI-LS's beam rotator (K-mirror) rotated the projection of the detector array onto the sky so that the y-axis of the detector had a position angle of $60\degr$ East of North. To cover the mapping area, the telescope was pointed in a $30\arcsec$ raster (tiling the blue array) also rotated by $60\degr$ resulting in maps aligned with the PDR so that the long edge of the maps run along the PDR, the short edge across it. Most maps are displayed like this with an arrow indicating north.

To subtract the atmospheric and most of the telescope background, the observations were chopped at 2\,Hz in the asymmetric chop mode with a {10\arcmin} chop throw. Ideally one would chop to the south-west, but mechanical limitations of the SOFIA secondary mirror do not allow a {10\arcmin} chop throw in all directions. At the time of the observations in 2016, a large chop throw to the south-west was not possible. Therefore, we chopped to the east as far as possible (position angles of $97\degr$ and $108\degr$ with chop throws between 9\arcmin\ and 10\arcmin). In 2019, a large chop towards the southwest was possible, and we chopped {9.83\arcmin} towards 240\degr. A comparison of the 2016 and 2019 data does not reveal any offsets introduced by the different chop angle, at least at the wavelengths, observed in 2016 and 2019, i.e. \oi63\um\ and \CO.

To subtract the residual telescope background, reference positions offset by (-800\arcsec,-400\arcsec) in RA-DEC relative to the map positions were observed with the same asymmetric chop in an $\rm A_1A_2BA_3A_4$ nod sequence (${\rm A}_i$ are different raster position while B is a reference position).  Details about the observing schemes used with FIFI-LS can be found in \cite{FIFILS_SPIE16}.

%\newlength{\RKdatelength}
%\settowidth{\RKdatelength}{2016-06-28}
%\newlength{\RKflightnolength}
%\settowidth{\RKflightnolength}{F\,999} %\begin{deluxetable}{lrp{\RKdatelength}p{\RKflightnolength}c}
\begin{deluxetable}{lrccc}
  % centered columns
  \tablecaption{List of Observations\label{tab:obs}}
  \tablehead{
    \colhead{Species} & \colhead{$\lambda$}&\colhead{Dates}&
    \colhead{Flight}&\colhead{Area}\\   
  & \colhead{[$\mu$m]} &        &           &
  \colhead{[$\arcmin\times\arcmin$]}}
  \startdata
  \oiii    & 51.815& 2016-06-28& F\,310 & $5\times3.5$\\
  \hline
  \niii    & 57.317& 2016-06-28 & F\,310 F\,314 F\,317 & $5\times3.5$\\[-.5ex]
  &&2016-07-03&&\\[-.5ex]
  &&2016-07-06&&\\
  \hline
  \oi      & 63.184& 2016-07-06 & F\,317 F\,572 & $5\times3.5$\\[-.5ex]
  &&2019-05-16&&\\
  \hline
  \oiii    & 88.356& 2016-07-03& F\,314 & $5\times3.5$\\
  \hline
  \oi      &145.525& 2016-06-28& F\,310 & $5.5\times4$\\
  \hline
  CO(17-16)&153.267& 2016-06-28 & F\,310 F\,314 F\,317& $5.5\times4$\\[-.5ex]
  &&2016-07-03&&\\[-.5ex]
  &&2016-07-06&&\\
  \hline
  \cii     &157.741& 2016-07-06& F\,317 & $5.5\times4$\\
  \hline
  CO(16-15)&162.812& 2016-07-06 & F\,317 F\,572 & $2.0\times2.7$\\[-.5ex]
  &&2019-05-16&&\\
  \hline
  CO(14-13)&185.999& 2016-07-03 & F\,314 F\,317 F\,572& $5.5\times4$\\[-.5ex]
  &&2016-07-06&&\\[-.5ex]
  &&2019-05-16&&\\
  \enddata
  \tablecomments{The table lists the observed species, rest wavelengths, flight dates, flight numbers, and map sizes.}
\end{deluxetable}

\begin{deluxetable*}{lr|D@{$\pm$}DDD@{$\pm$}DD|r}
%\begin{deluxetable}{lr|D@{$\pm$}DD|r}
    \tablecaption{Peak fluxes and noise levels of the line intensity maps\label{tab:values}}
  \tablehead{
    \colhead{Transition} & \colhead{T$_{\mathrm{int}}$}&    \multicolumn{4}{|c}{peak}&\multicolumn2c{rms}&\multicolumn{4}{c}{peak sm.}&\multicolumn{2}{c|}{rms sm.}&
\colhead{R}\\   
& 
\colhead{[sec]}&
\multicolumn{12}{|c|}{[$\mathrm{\frac{10^{-3}erg}{s\,cm^2sr}}$]}&
%\multicolumn{6}{|c|}{[$\mathrm{\frac{10^{-3}erg}{s\,cm^2sr}}$]}&
\colhead{$\frac{\lambda}{\Delta\lambda}$}
}
  \decimals
  \startdata
  \oiii52\um    &  717 &261.    &37.    &9.1   &213.  &30.   &6.4   &
  1020\\
  \hline
  \niii57\um    &  727 & 44.6   & 6.3   &3.3   & 38.6 & 5.5  &1.6   &1100\\
  \hline
  \oi63\um      &  1976& 25.9   & 5.3   &1.5   & 17.1 & 3.3  &0.81  &
  1300\\
  \hline
  \oiii88\um    &   717& 91.    &13.    &3.0   & 80.  &11.   &3.0   & 630\\
  \hline
  \oi146\um     &   717&  4.27  & 0.97  &0.19  & 4.05 & 0.91 &0.16  &
  1100\\
  \hline
  CO(17-16)     &   727&  0.34  & 0.23  &0.064 & 0.30 & 0.19 &0.033 &
  1000\\
  \hline
  \cii          &   717&  4.71  & 0.67  &0.32  & 4.59 & 0.65 &0.30  &
  1100\\
  \hline
  CO(16-15)     &  1321&  0.53  & 0.11  &0.056 & 0.52 & 0.10 &0.046 &
  1100 \\
  \hline
  CO(14-13)     &  1670&  0.679 & 0.097 &0.064 &\multicolumn{6}{c|}{N/A}  &
  1600\\
  \hline
  \ifir$\times10^{-3}$&N/A &\multicolumn{6}{c}{N/A}  &18.     &7.2  &1.3   &N/A\\
  \enddata
\tablecomments{For each transition: total on-source integration time, peak flux with its uncertainty and the median uncertainty over the map (uncertainties as described in Sect.~\ref{ssec:dataredu}) for the original and smoothed (sm.) map, and spectral resolution used in the line fitting.}
\end{deluxetable*}

\begin{figure}
  \includegraphics[width=\linewidth,viewport=35 8 485 342]{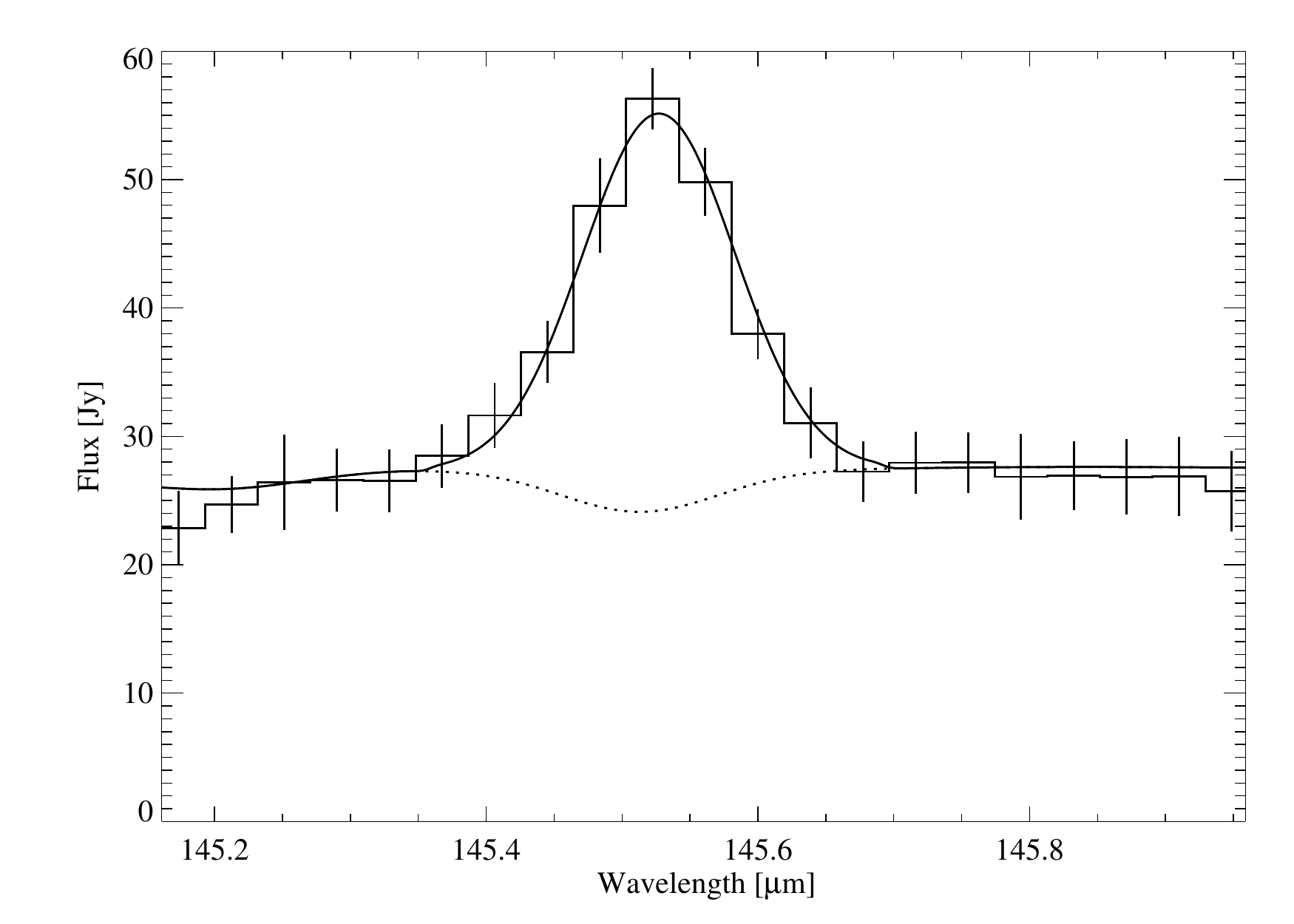}
  \caption{The observed \oi146\um\ line and continuum (histogram with error bars) at a representative position in the PDR (pos.~2 in Fig.~\ref{fig:lines}) fitted with a Gaussian (solid line) and an ATRAN baseline (dotted). See also Sect.~\ref{ssec:dataredu}.}
  \label{fig:spectrum}
\end{figure}

\begin{figure*}
  \includegraphics[height=0.407\linewidth,viewport=36 25 415 325]{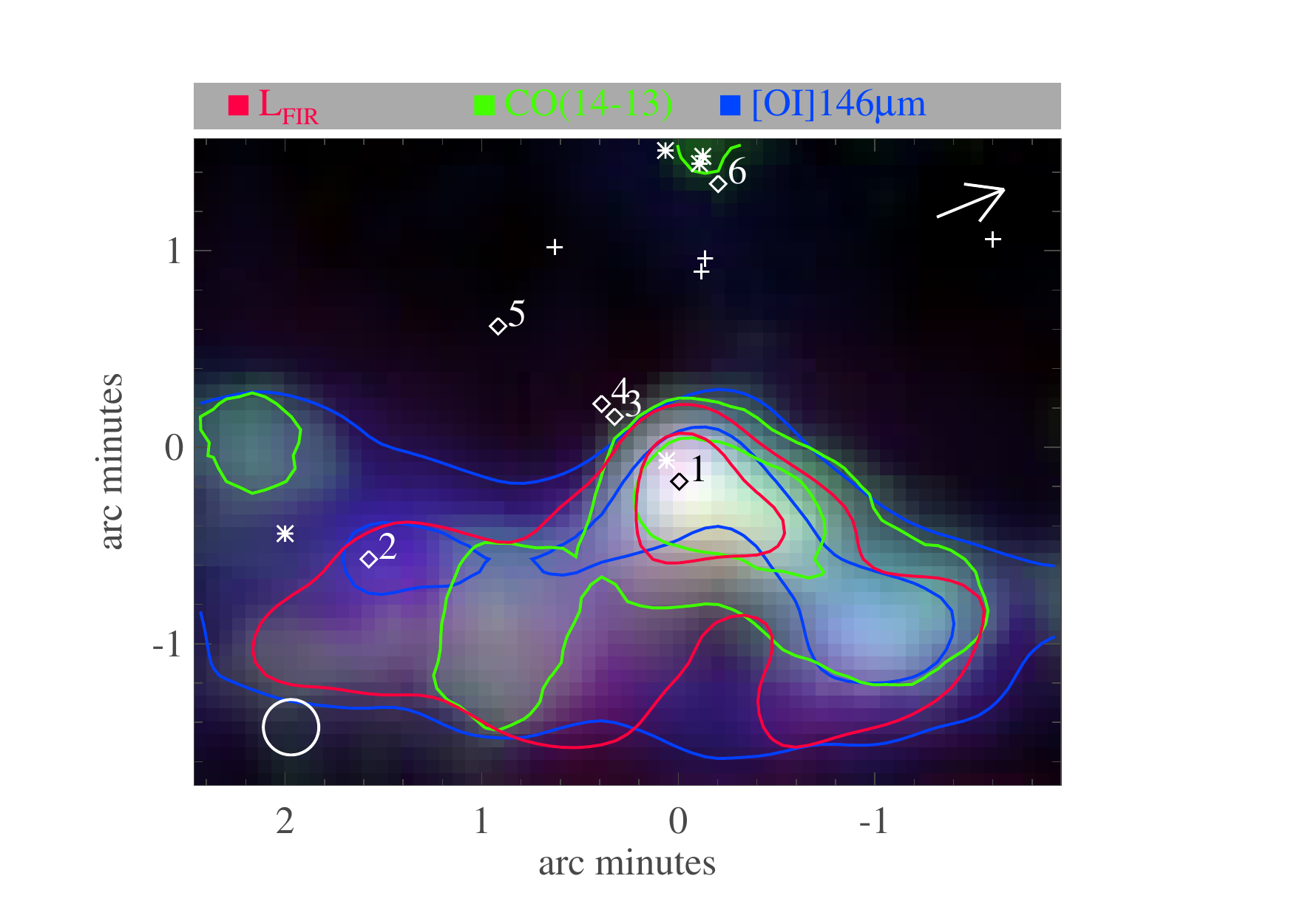}
  \includegraphics[height=0.407\linewidth,viewport=58 25 415 325]{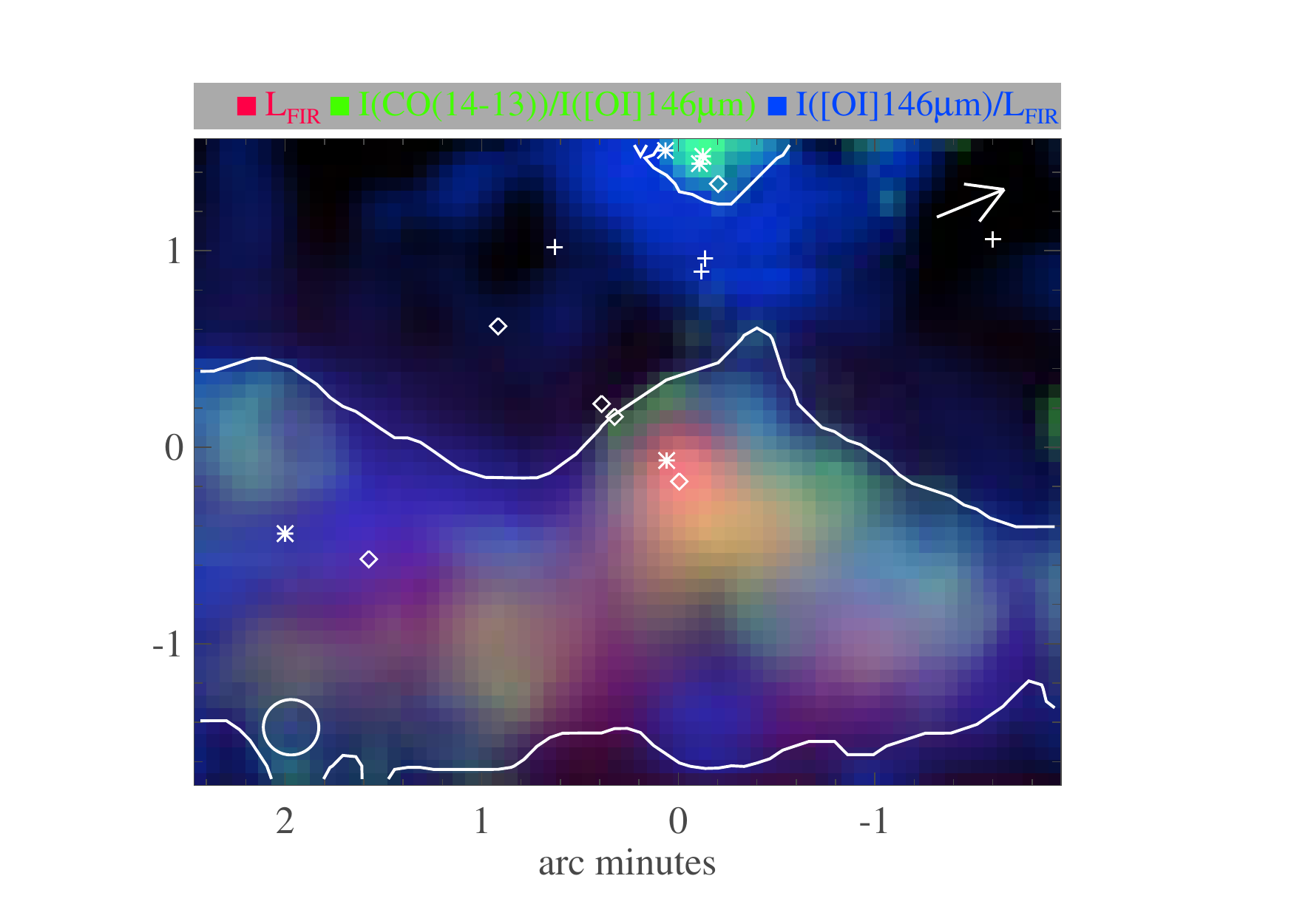}
  \caption{The PDR model input: left - the three observed intensities maps used to calculate the input quantities for the PDRT model; right - the two ratio maps and the \ifir\ map used in the PDR modeling.  The colors start to fade from a signal-to-noise ratio of 3 and reach black at a ratio of 1.  The contours on the left are at $1/3$ and $2/3$ of the peak fluxes (Tab.~\ref{tab:values}). The contour on the right traces the density jump as determined by the model (see Sec.~\ref{sssec:density-uv}). The diamonds mark the positions referenced in Fig.~\ref{fig:chi2}. The stars and crosses mark the O-stars as in Fig.~\ref{fig:fs_lines}. The arrows point north and the circles indicates the common spatial resolution to which all maps used in the PDR model were smoothed.}
\label{fig:lines}
\end{figure*}

\subsection{Data Reduction}
\label{ssec:dataredu}

The FIFI-LS data reduction pipeline \citep{Vacca2020} was used to reduce the raw data. The final data products of the pipeline (level 4) are data cubes with the observed flux regridded on an oversampled, regular, three dimensional $(x,y,\lambda)$-grid with the $x$- and $y$-axes usually aligned with equatorial coordinates. However, for this data set, we kept the $x$- and $y$-axes aligned with the array and raster map axes, which were rotated by $60\degr$ relative to the equatorial coordinates. 

The statistical uncertainties were estimated using an algorithm developed by one of us (D. F.) analogous to the way uncertainties were estimated for the PACS-spectrometer on board Herschel\footnote{The Python implementation of the algorithm can be found here under Cubik: \texttt{https://github.com/darioflute/fifipy}}. The uncertainty for each pixel in the data cube were based on the variance of all individual measurements in the spatial (within two times the full-width-half-maximum (FWHM) of the beam and within three times the FWHM for $\lambda<65\um$) and spectral (1/4  of FWHM of the line spread function) vicinity of that pixel. The error is computed by weighting the contributions by the distance in space and wavelength.

The data is flux-calibrated in the pipeline processing using calibration factors derived as described by \cite{Fischer2018} achieving an absolute calibration of better than 15\%.  To not underestimate such systematic uncertainties, a relative uncertainty of 10\% is added in quadrature to the propagated statistical uncertainties.  The uncertainty for the atmospheric correction  for the \edit1{{\oi146}\um} line is higher (possibly 20\%) as the line is close to a narrow but deep telluric absorption feature (see Fig.~\ref{fig:spectrum}).

To obtain the integrated line fluxes and continuum flux densities, we fit all spectra with a Gaussian and a baseline. The only free variable for the Gaussian was the line flux. Line center and width were fixed to the transition's wavelength and the instruments resolution; the observed spectra did not exhibit any wavelength shifts or resolved line profiles. The baseline was a constant flux density $F_\nu$ multiplied with the atmospheric transmission modelled with ATRAN \citep{ATRAN} convolved to the instrument's spectral resolution.

The parameters for the ATRAN-model are altitude, zenith angle, and precipitable water vapor (PWV). While altitude and zenith angle are known from the observations, the PWV values are not directly available. Since the {\oi63\um} and the \CO\ lines are on the wings of broad telluric water absorption features and the spectra have strong continua, we  were able to fit the atmospheric model to the spectra by letting PWV as a free parameter. We obtained values between 2 and {3\um} of PWV for these flights which are consistent with those derived from satellite data \citep{iserlohePWV} for the 2016 data as well as the FIFI-LS measurements taken directly before and after the data acquisition in 2019 \citep{fischerPWV}. Thus we used {2.5\um} PWV for all observations. The only remaining free parameter for the baseline fit was the constant continuum flux density. Having determined the PWV for the observations allowed  the correction of the observed line and continuum fluxes for atmospheric absorption.

An example for such a line and baseline fit is shown in Fig.~\ref{fig:spectrum}. Shown is the \oi146\um\ at a representative PDR position (position 2 of the discussion in Sect.~\ref{sec:analysis} and Fig.~\ref{fig:lines}). The histogram with the errorbars shows the observed spectrum with its uncertainties computed as previously discussed. 
The solid lines show the fit (Gaussian emission line plus baseline) while the dashed line corresponds to the baseline. The dip in the baseline stems from a narrow atmospheric absorption feature broadened to the instrument's spectral resolution.
 
Some data sets presented specific problems which were addressed in our analysis as explained in the following.

The continuum of the \oiii52\um\ observations was not used to estimate the far-infrared intensity (see Sect.~\ref{sec:analysis}) as the continuum map showed strong artifacts stemming from bad data at some map and especially reference positions. As the data from the reference position is used to subtract the sky emission at several map positions, a relatively large part of the map is affected. The \oiii52\um\ line intensity map is not affected as the bad reference data introduced only an offset in the continuum levels.

The \cii\ line intensity map showed clear indication of \cii\ emission in the off-beam for the chop-pairs especially in the northern corner of the mapped area. The \cii\ emission in the off-beam was stronger than in the on-beam so that the resulting spectra showed the \cii\ line in absorption rather than in emission there. To obtain a \cii\ map, the data was reduced in the same way as the other data sets except that the data from the off-beam from each chop-pair was not subtracted, but ignored. Only the on-beams of the reference positions were subtracted from the on-beams in the source positions. That allowed to recover the \cii\ flux\footnote{In 2020, a total power mode, which does not chop but only nods the telescope, has been commissioned and produced well calibrated data}. Similarly, a map of the \cii\ emission in the off-beams relative to off-beams at the reference position can be made. That showed that the off-beams were free of \cii\ emission in \edit1{all parts} of the map except in the northern corner of the mapped area, where the off-beams saw up to 1.5\merg\ of \cii\ in a ridge roughly at a declination of $-16\degr11\arcmin$ seen from RA $18^\mathrm{h}20^\mathrm{m}55^\mathrm{s}$ to $18^\mathrm{h}21^\mathrm{m}10^\mathrm{s}$ matching the far eastern end of M17 North (just outside of Fig.~\ref{fig:m17-sw}). 

The absolute continuum level in this virtually un-chopped observation was not used, as the atmospheric background subtraction is much less certain due to the about 2 orders of magnitude slower nod frequency compared to the chopping frequency. The 158\um\ continuum map was obtained from the normally reduced \cii\ observations. At no wavelengths are there any indications of significant continuum emission in the off-beams.

Four of the maps are displayed in Fig.~\ref{fig:fs_lines} as colored overlays at their original resolutions. Before calculating intensity ratios for the PDR modeling, each map is convolved with a Gaussian to match the spatial resolution of the longest wavelength map. The individual line maps at their original resolution and the smoothed version are shown in Appendix~\ref{sec:maps} together with a FIR intensity (\ifir) map. We created the \ifir\ map by creating spectral energy distributions (SEDs) for each map pixel in our continuum maps and available Herschel continuum maps and then integrated the SEDs to obtain the \ifir\ at each position. We estimate a 40\% uncertainty for this map. The FIR continuum is shown together with the line maps that enter the PDR model, {\oi146\um} and \CO, in the left panel of Fig~\ref{fig:lines}. For the convolved maps, the uncertainties are propagated taking the correlation in the original maps into account \citep{Klein2021_RNAAS}. We assume that the correlation between pixels in the original maps is on spatial scales of the respective beam sizes.

\subsection{Flux Cross-Calibration}
\label{sec:flux-calibration}

Comparing directly to the Kuiper Airborne Observatory (KAO) observations of M17-SW by \cite{Stutzki} and \cite{Meixner92}, we find a general agreement with our SOFIA/FIFI-LS observations of fine-structure lines. For example, along the two scans done by \cite{Stutzki}, the \cii-emission smoothed at the same resolution as the KAO data also peaks at \mbox{$3.5$\merg} and is about 0.5\,pc wide. However, we find the location of the \cii-peak about 0.5\arcmin\ further southwest along the scan.

The FIFI-LS \cii-map matches the integrated intensity \cii-map obtained with SOFIA/GREAT \citep{Perez2012} very well in shape, position, and relative intensity of the emission. The absolute flux calibration of our \cii-data seems to be about 20\% lower than the SOFIA/GREAT observations, which is within the combined uncertainties of both observations.

\begin{figure}
  \includegraphics[height=.475\linewidth,viewport=35 7 358 308]{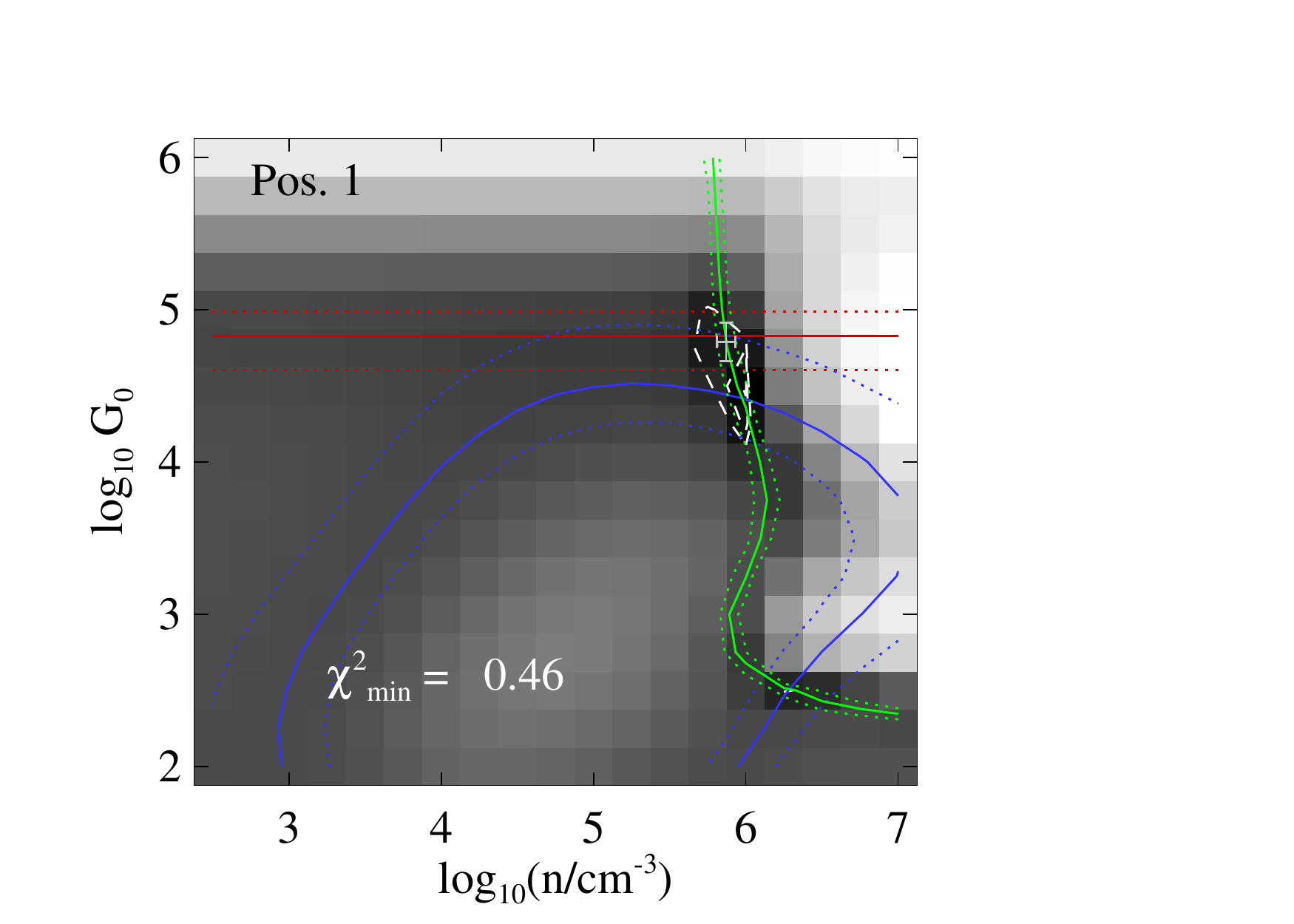}
  \includegraphics[height=.475\linewidth,viewport=55 7 358 308]{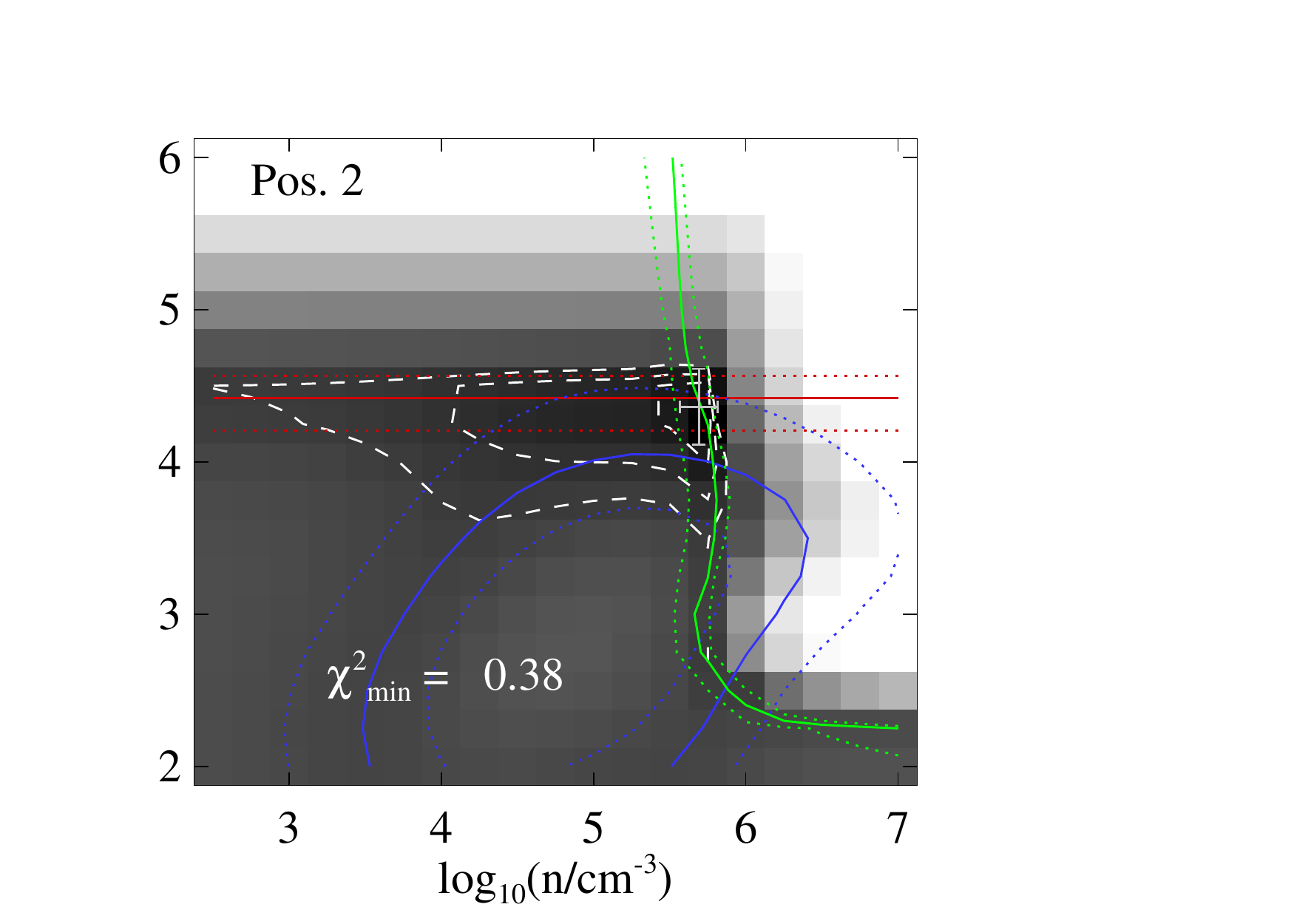}\\
  \includegraphics[height=.475\linewidth,viewport=35 7 358 308]{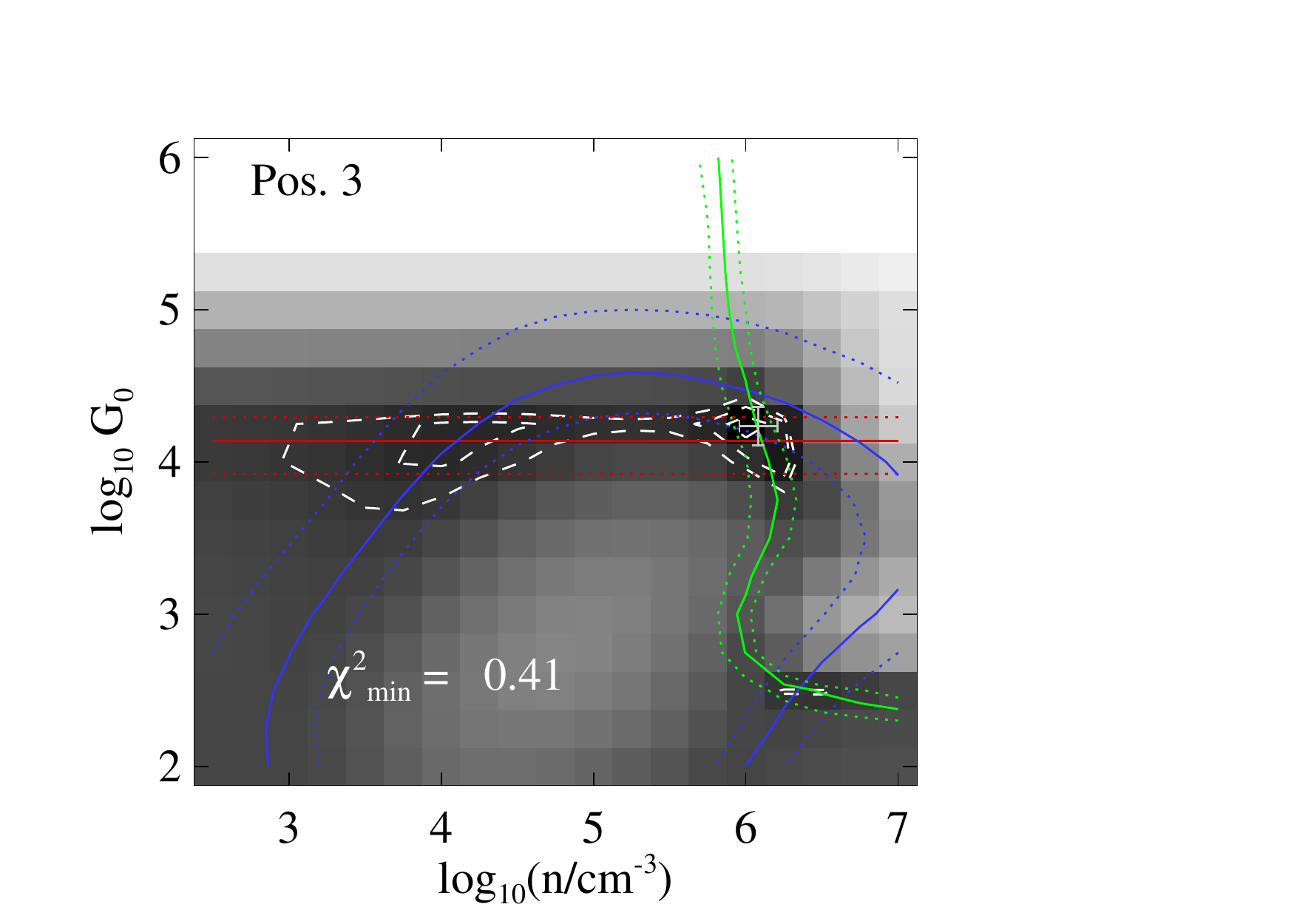}
  \includegraphics[height=.475\linewidth,viewport=55 7 358 308]{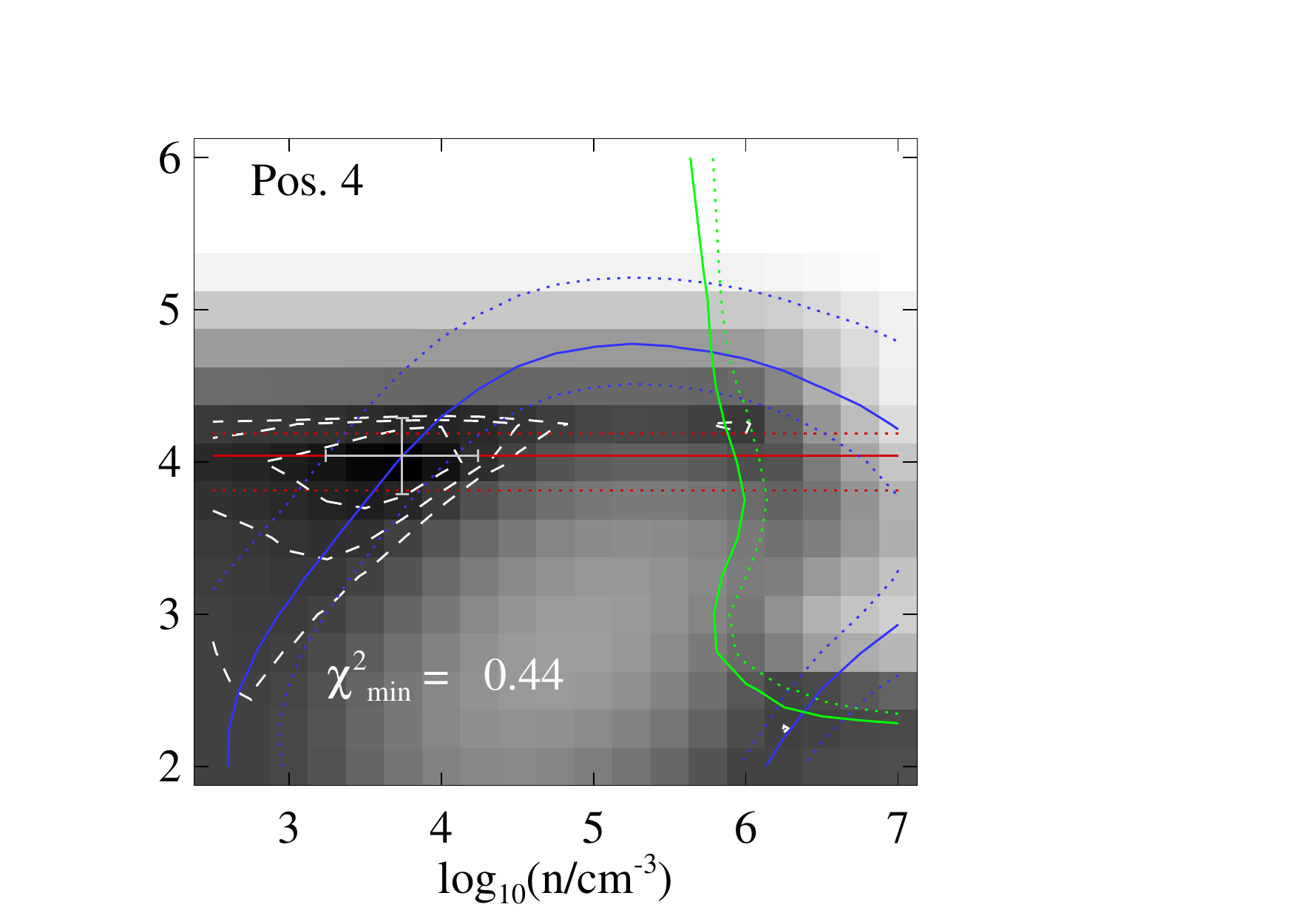}\\
  \includegraphics[height=.475\linewidth,viewport=35 7 358 308]{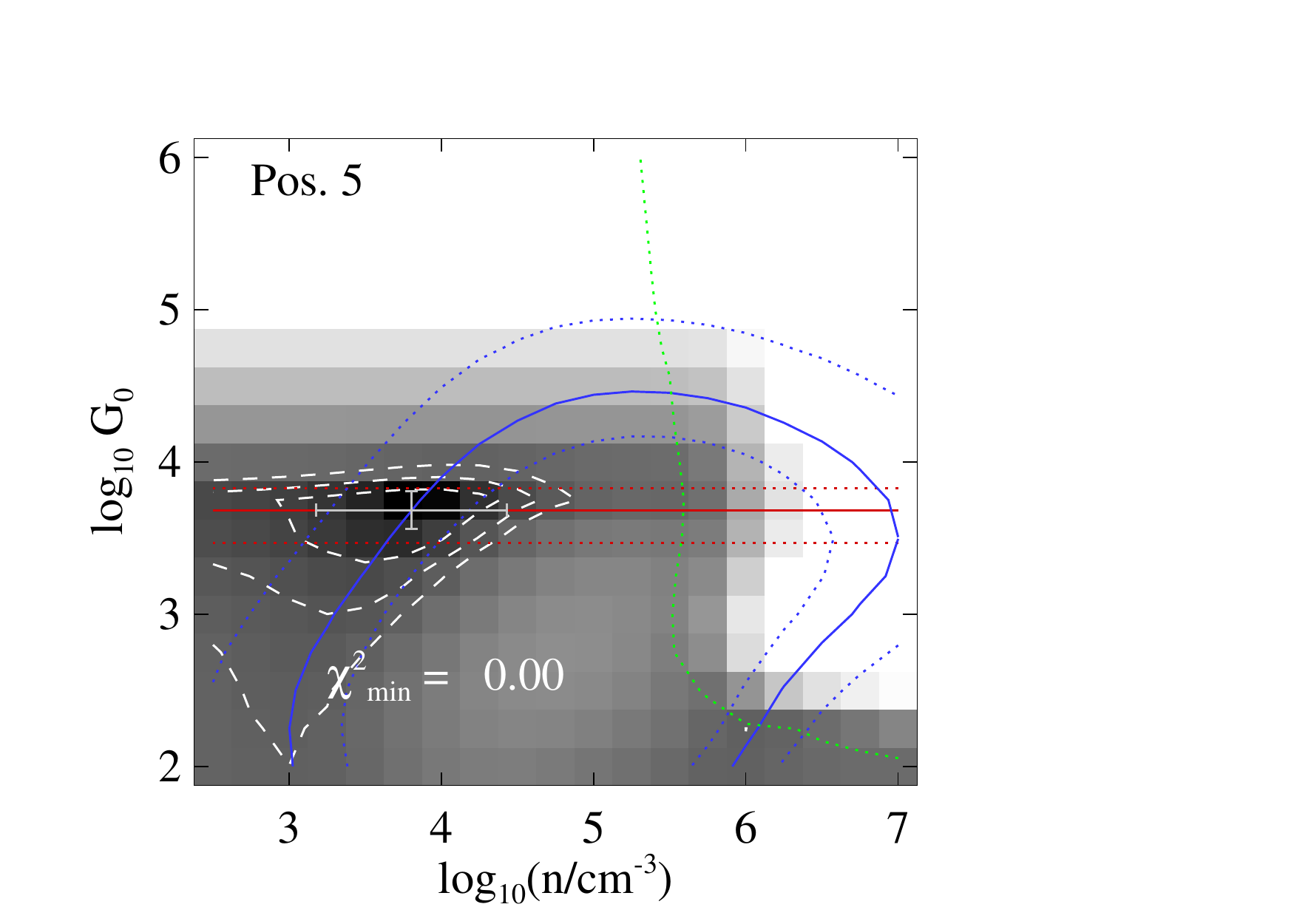}
  \includegraphics[height=.475\linewidth,viewport=55 7 358 308]{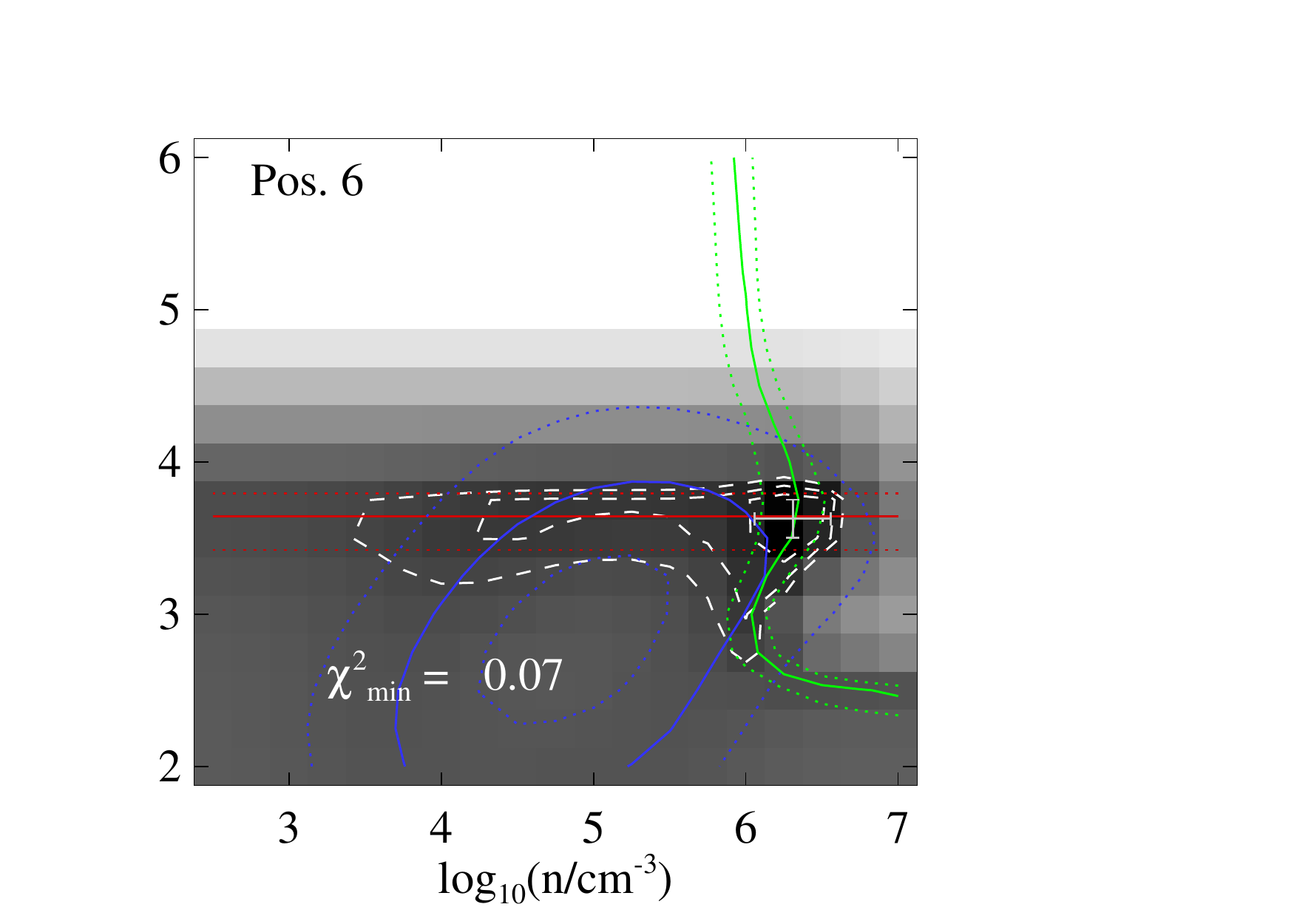}
  \caption{Reduced $\chi^2$ (logarithmic greyscale up to $10^3$) in $n$-{\GUV}-plane for the six positions marked in Fig.~\ref{fig:lines}.  In each panel, the grey errorbars mark the minimum $\chi^2$ location with dashed contours marking the increase of $\chi^2$ by one, two, and three, respectively. The minimum $\chi^2$-value is given for each panel. The colored solid lines mark the {\nG}-pairs predicting the observed line ratios and FIR intensity (red: \ifir, green: I(\CO)/I(\oi146\um), and blue: I(\oi146\um)/\ifir). Dotted lines indicate 1$\sigma$ uncertainty of the observations. Note that for pos.~5, there is no solid green line because the observed \CO\ to \oi146\um\ line ratio is too low. For pos.~4, only the dotted line for the lower uncertainty limit vanishes.}
  \label{fig:chi2}
\end{figure}

\begin{figure*}
  \includegraphics[height=0.368\linewidth,viewport=36 19 460 327]{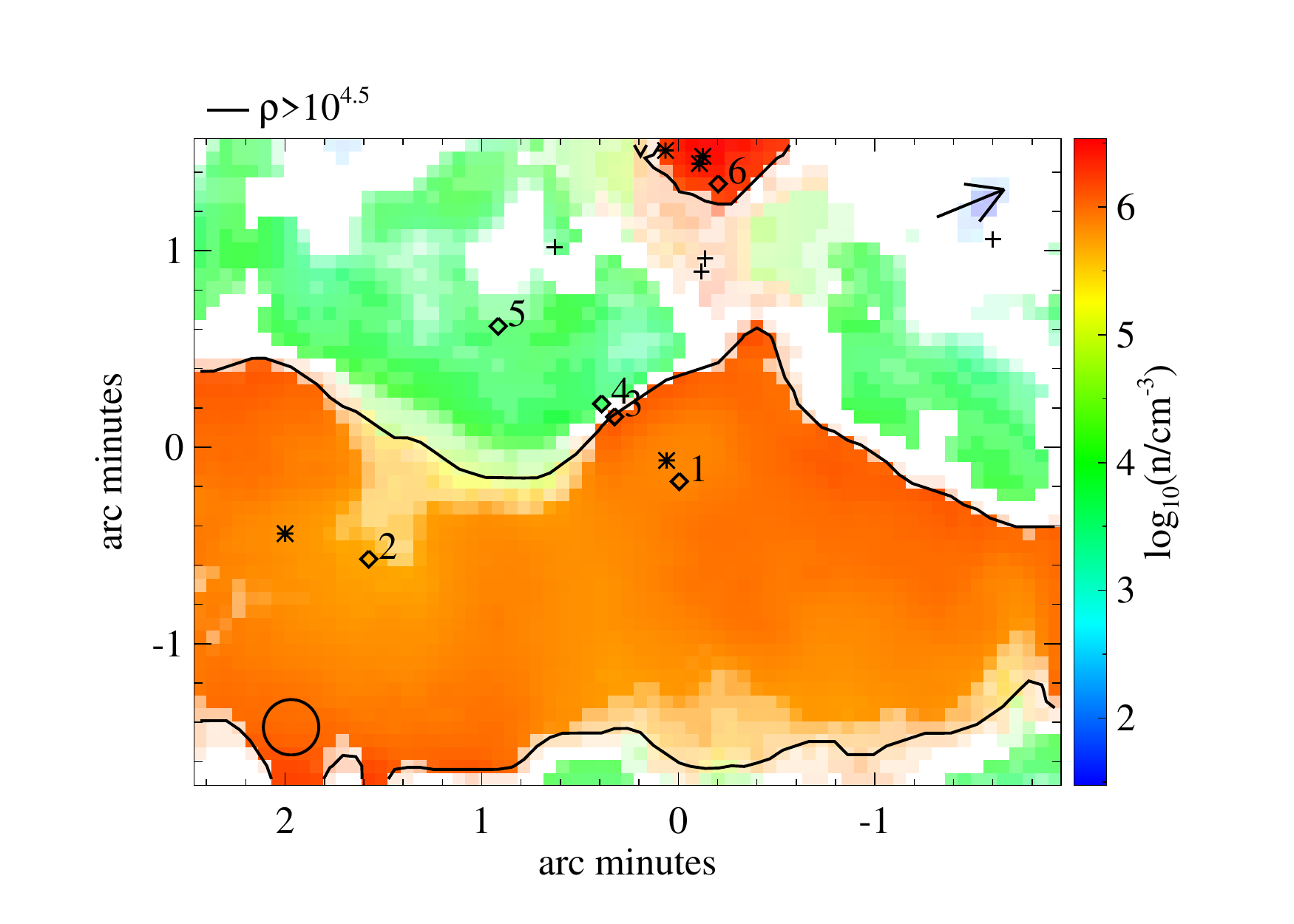}
%  \hfill
  \includegraphics[height=0.368\linewidth,viewport=58 19 470 327]{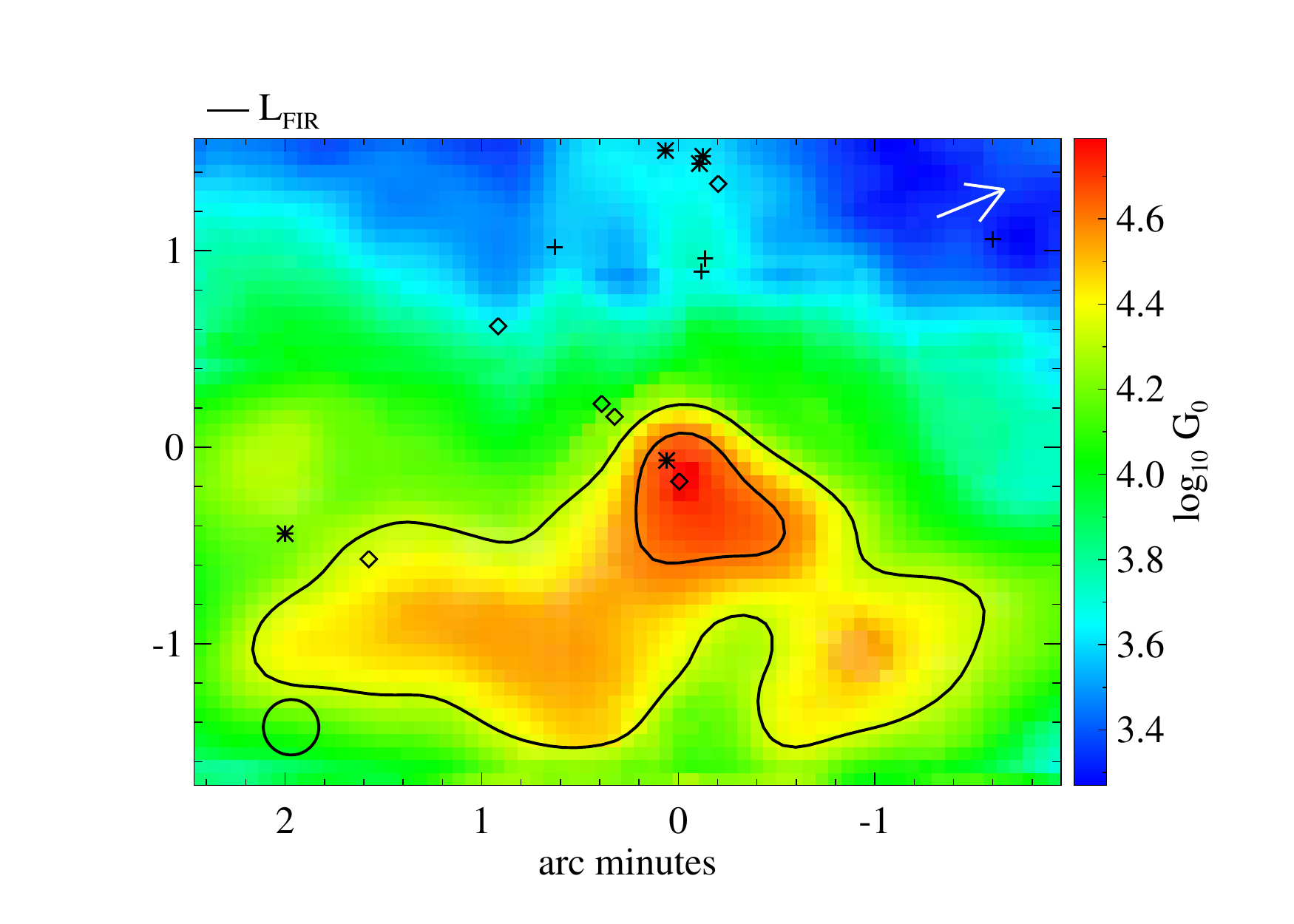}
  \caption{The modeling results: left - the H-density and right - the UV radiation field. The color bars show the color scale used for values with an uncertainty less than 0.2\,dex. The color fades with increasing uncertainties becoming white for uncertainties larger than 1.25\,dex. Diamonds, circles, and arrows are as in Fig.~\ref{fig:lines}. The contour in the density plot traces the density jump (Sec.~\ref{sssec:density-uv}). The contours in the UV field plot are the same as in Fig.~\ref{fig:lines} for \ifir.}
  \label{fig:results}
\end{figure*}

\begin{figure*}
  \includegraphics[height=.369\linewidth,viewport=36 19 460 327]{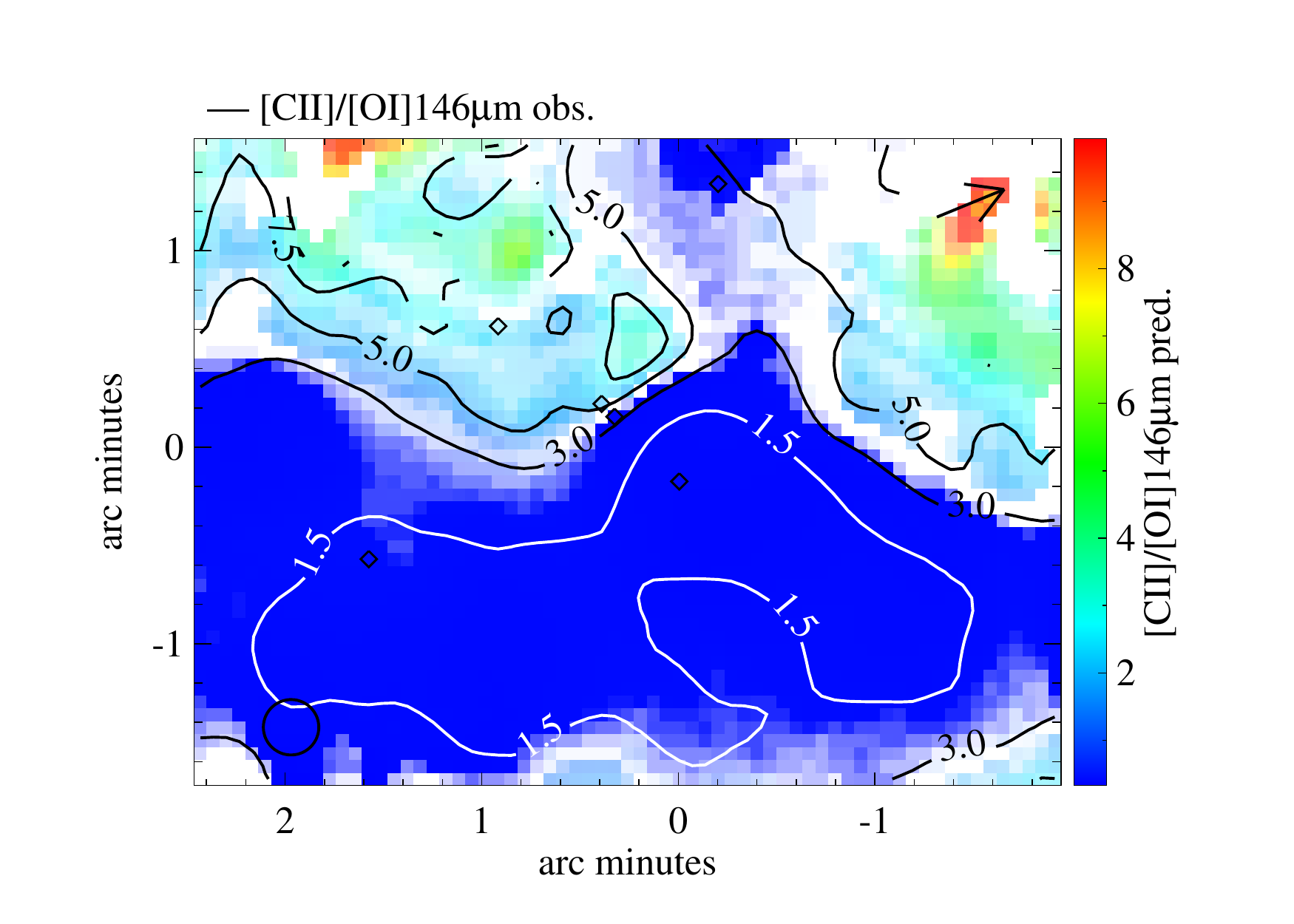}
  \includegraphics[height=.369\linewidth,viewport=58 19 468 327]{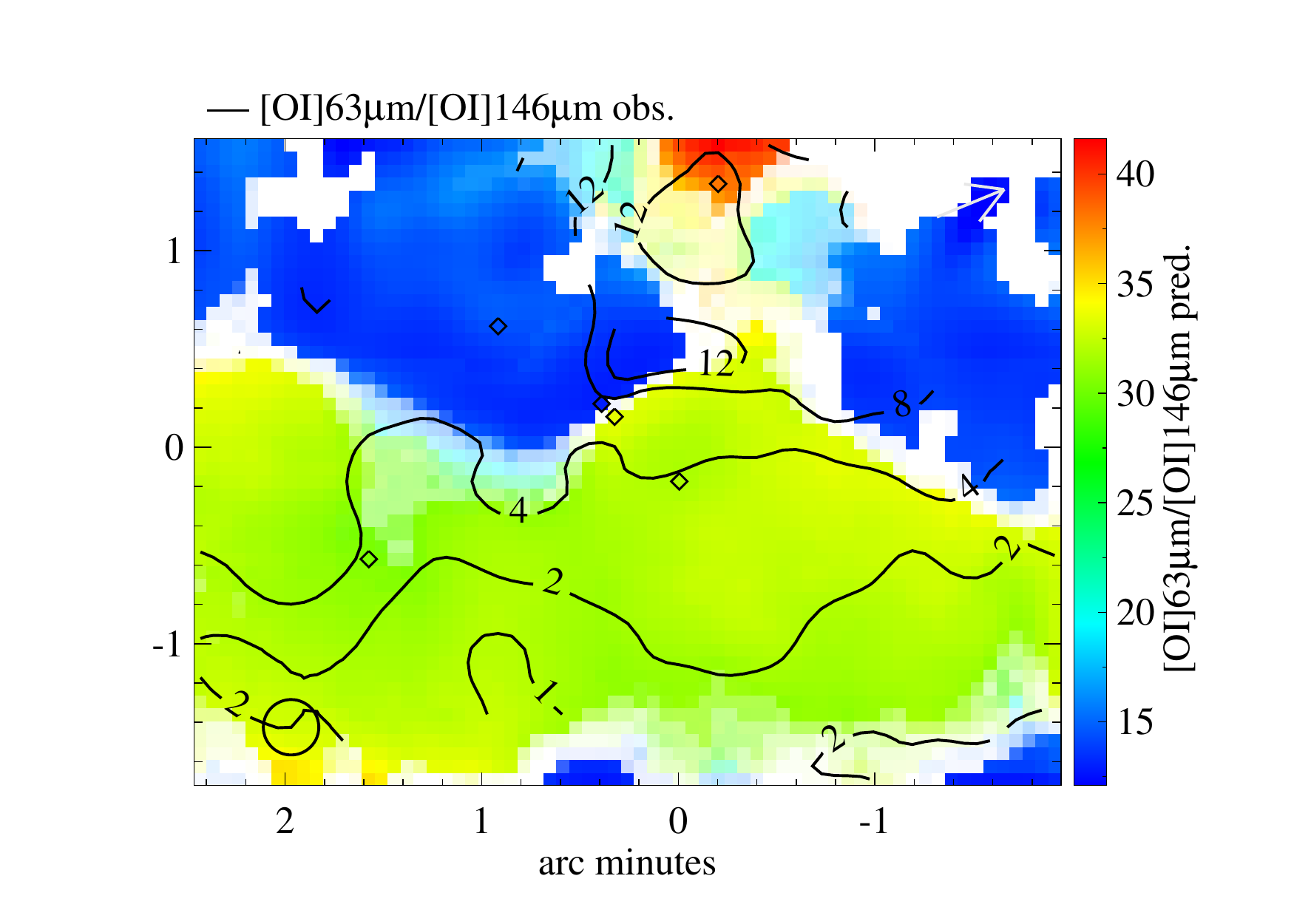}
  \caption{The \cii\ and \oi63\um\ to \oi146\um\ line ratios: predictions vs  observations; the color scale represents the predicted line ratio (colors fading from an SNR of 5 to 1 (for \cii/\oi146\um) and 2 (for \oi63\um/\oi1246\um). The contours trace the observed line ratio (SNR $>2$). Symbols are like in Fig.~\ref{fig:results}.}
  \label{fig:ci_oi_ratios}
\end{figure*}

% \begin{figure}
%   \includegraphics[width=\linewidth,viewport=48 23 380 308]{inspect_oicii_pred-eps-converted-to.pdf}  
%   \caption{ \oi63\um\ to \cii\ line ratio predictions vs observations in the \nG-plane.}
%   \label{fig:oicii_ng}
% \end{figure}

\section{PDR Analysis}
\label{sec:analysis}

The PDR Tool Box \citep[PDRT]{Kaufman2006_PDRT,Pound08_PDRT}%
\footnote{http://dustem.astro.umd.edu/ The PDR Tool Box (PDRT) was significantly revised in August 2020, when the analysis presented here had already been finished. In terms of the new PDRT, we are using the wk2006 models.} predicts ratios and some intensities of the main cooling lines of a PDR depending on the density of hydrogen nuclei (H-density), $n$, and the strength of the UV radiation field, {\GUV}. In addition, the far infrared continuum is predicted assuming that all the UV radiation is absorbed and re-emitted in the far infrared. The comparison between observations and predictions provides us with estimates of $n$ and {\GUV} for each position in the mapped area.

The PDRT is a one-dimensional face-on model. The model predicts line intensities or ratios only as a function of the hydrogen density and UV intensity, but not as a function of depth into the cloud.  Looking at the M17-SW PDR mostly edge-on, we should see optical depth effects going from the \hii\ region into the molecular cloud for lines subject to self-absorption.
Therefore, we only compare the optically thin {\oi146\um} and \CO\ lines and \ifir\ to the PDRT model. The two lines have a lower energy levels of 228\,K and 503\,K, respectively, which can only be sparsely populated in the molecular cloud and, thus, these lines cannot show significant self-absorption. The FIR continuum will also be mostly optically thin. Since the optically thin emission can escape in any direction, the observed intensities should not depend much on the exact geometry of the PDR like the inclination angle of the PDR surface nor the clumpiness of the medium.  Taking the line ratio can further eliminate any systematic effects.

For the analysis, the maps were smoothed to the spatial resolution of the \CO\ map, which has the largest beam size (longest wavelength: 186\um, beam FWHM: 17\arcsec). From these three optically thin quantities, we computed two line ratios,  I(\CO)/I(\oi146\um) and I(\oi146\um)/\ifir. These ratios and the FIR intensity were compared to the PDRT predictions to determine the H-density, $n$, and UV radiation field, {\GUV}. Fig.~\ref{fig:lines} shows the line ratio maps and the \ifir\ map on the right and on the left the line intensity maps (and \ifir) from which the ratios are derived.

While the {\cii} line is a significant cooling line for PDRs, C$^+$ is ubiquitous in M17 as previous studies (see references in Sec.~\ref{sec:intro}) and our inadvertent detection of \cii\ in some off-beams shows, which indicates that the observed \cii\ emission may not only be associated with the PDR in M17-SW.  Furthermore, the \cii\ and \oi63{\um} lines exhibit self-absorption by a colder foreground material or optical depth effects in the PDR as already pointed out by \cite{Perez2012} and \cite{Perez_Ringberg2017}. Furthermore, \cite{Perez2015_atomicgas} show that significant fractions of the \cii-emission come from the ionized, atomic, and molecular gas phase. Therefore, we do not use the \cii\ and the \oi63\um\ line as input parameters for the PDRT modeling, but we compare the predictions of the PDRT-model for these lines to the observations in Sect.~\ref{ssec:predictions}.

For each pixel $\vec{x}$, we determine the absolute minimum for the reduced $\chi^2$ 
\begin{equation*}
  \chi^2(\vec{x},n,G_{\mathrm UV})=\frac{1}{2}
  \sum_{i=1}^3\left(\frac{O_i(\vec{x})-M_i(n, G_{\mathrm UV})}
    {\sigma_i(\vec{x})}\right)^2
\end{equation*}
with $O_i$ being the three observed quantities, {$M_i(n,$}\GUV{$)$} the model predictions for these quantities, and $\sigma_i$ the uncertainties (one standard deviation) for the observed quantities. The model predictions {$M_i(n,$}\GUV{$)$} are interpolated from a logarithmically sampled ($n$,\GUV)-grid provided by the PDRT. The factor $1/2$ is due to the three degrees of freedom. The reduced $\chi^2$-distributions are displayed in Fig.~\ref{fig:chi2} for six positions chosen either as representative positions or to illustrate specific points in the discussion.  The positions are marked with diamonds in most figures with maps. The minimum $\chi^2$-values at these six positions is labeled in these plots. The values of the reduced $\chi^2$ minima over the whole map range from 0 to 1.3. Thus, fits are acceptable for all positions. The ($n$,\GUV)-pair where $\chi^2(\vec{x},n,G_{\mathrm UV})$ reaches its absolute minimum for a given map pixel $\vec{x}$ are the H-density and UV-radiation field at that position.

The derived H-densities (Fig.~\ref{fig:results}, left) form a bimodal distribution with peaks at $10^{3.8}\qcm$ and $10^{5.9}\qcm$ and a distinct jump in density. The UV map (Fig~\ref{fig:results}, right) shows UV intensities of up to $10^{4.7}\,G_0$ and a median of $10^{4.1}\,G_0$, with $G_0$ being the Habing field (0.12\merg).

The uncertainties of the derived  {\nG}-pairs varies significantly from position to position as the dashed contours for the 1-, 2-, and $3\sigma$-neighborhoods around the minimum indicate in Fig.~\ref{fig:chi2}. By looking up the extrema for $\log_{10}(n)$ and $\log_{10}$(\GUV) in the $1\sigma$-neighborhoods around the absolute $\chi^2$-minimum and taking the difference, we derived numerical uncertainties $\log_{10}(n)$ and $\log_{10}(${\GUV}$)$ (displayed as errorbars in Fig.~\ref{fig:chi2}). Note that the best fitting values for $\log_{10}n$ and $\log_{10}(${\GUV}$)$ are not necessarily near the mid-point between the extrema. Even in extreme examples such as positions 2 and 6 in Fig.~\ref{fig:chi2} this method allows us to derive meaningful uncertainties for H-density, UV intensity, and predicted line ratios. 

Having established the H-density and UV intensity for each map pixel, we compared the predictions of PDRT for line ratios including the optically thick lines {\oi63\um} and {\cii} to our observations.
The predicted and observed line intensities relative to the \oi146\um\ intensity are displayed in Fig.~\ref{fig:ci_oi_ratios} and discussed in Sec.~\ref{ssec:predictions}.

\section{Discussion}
\label{sec:discussion}

The FIFI-LS observations show a clear layering indicating an edge-on PDR. There is a separation between \oiii\ and \niii\ lines coming from the \hii\ region (discussed in a forthcoming paper) and \oi\ lines emitted from the PDR as can be seen in Figs.~\ref{fig:m17-sw} and \ref{fig:fs_lines}, but there is also a noticeable overlap between the regions of ionized and atomic oxygen emission. In the following sections, we discuss the results of the PDR analysis for the structure of M17-SW.

\begin{figure*}[tb]
  \includegraphics[height=.379\linewidth,viewport=44 26 429 312]{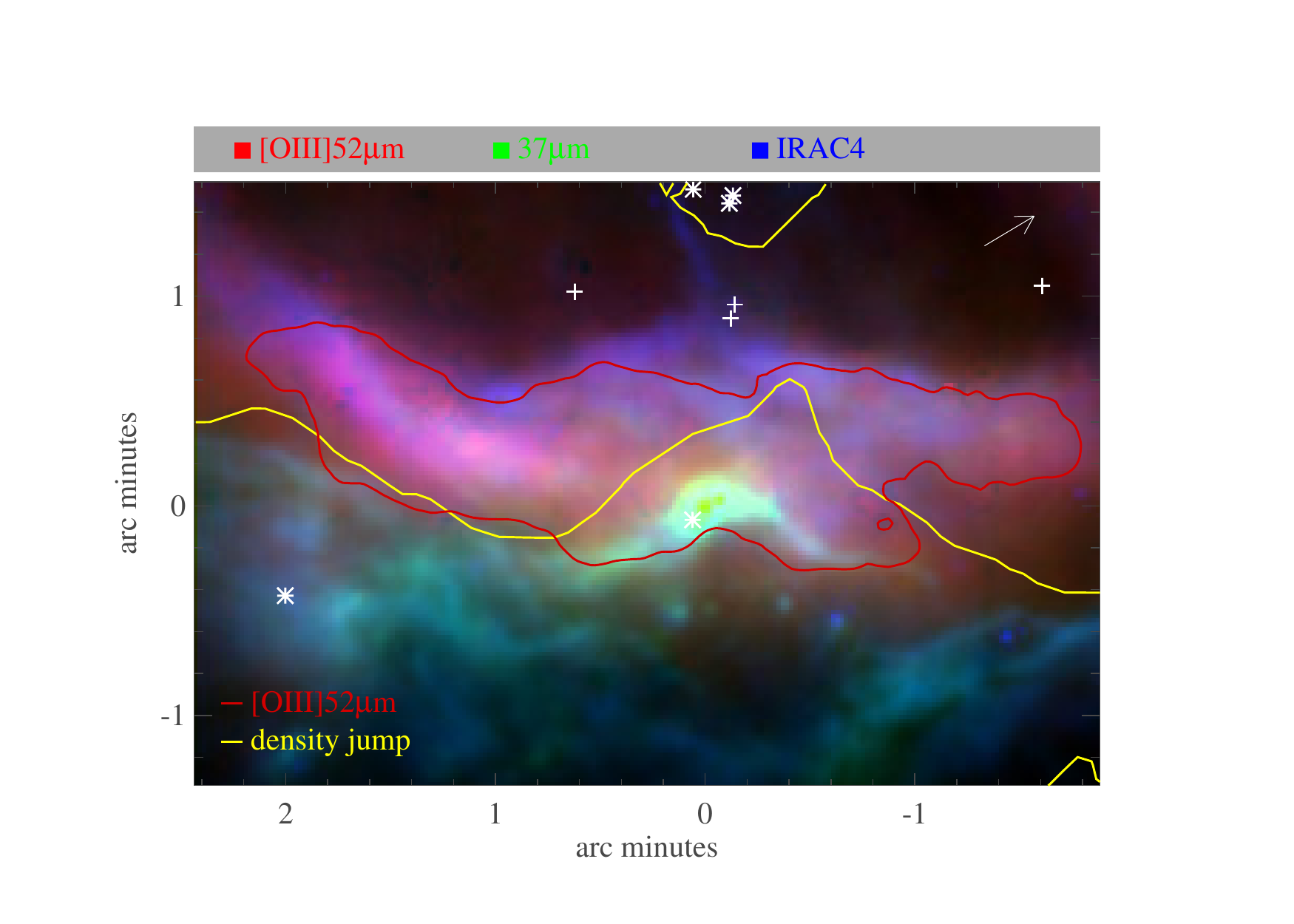}
  \includegraphics[height=.379\linewidth,viewport=60 26 429 312]{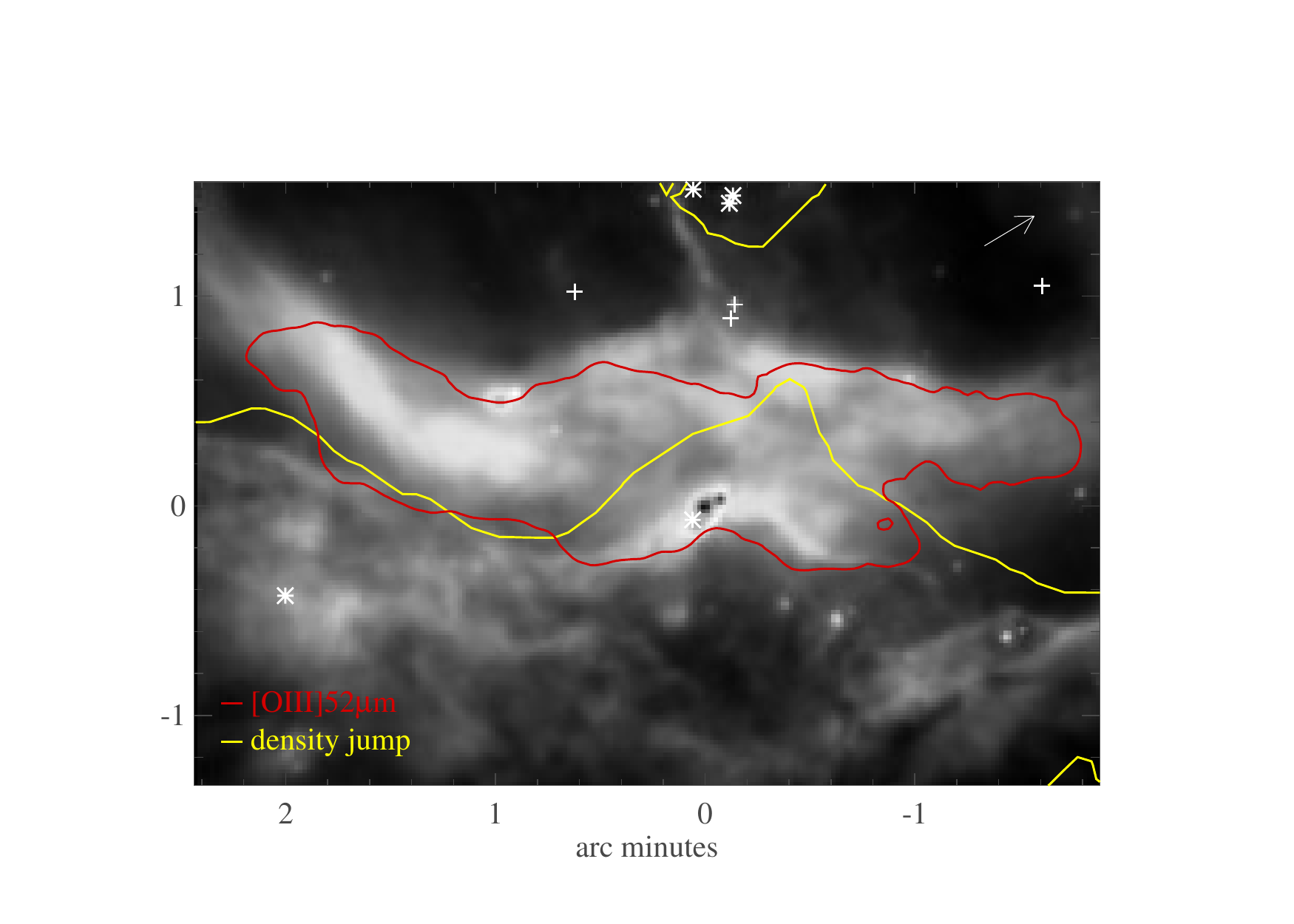}
  \caption{Illustrating the location of the front: Left - IRAC Band 4 (blue), FORCAST 37\um\ (green), \oiii52\um\ (red); Right - only the IRAC Band 4 band for clarity (saturated near M17-UC1); the red contour at 50\% of the peak intensity of the \oiii52\um\ line (Tbl.~\ref{tab:values}). The yellow contour traces the density jump. The stars and crosses are the locations of O-stars as in the figures before.}
  \label{fig:structure}
\end{figure*}

\subsection{ Gas Density and UV radiation}
\label{sssec:density-uv}

The PDRT assumes that all UV radiation is absorbed and remitted in the FIR. This is the reason why the derived UV-field closely follows the FIR intensity as indicated by the contours in the right panel of Fig.~\ref{fig:results}. It looks as if the contours represent the UV-field but actually they represent the FIR intensity.

The H-density map derived by the PDRT shows a distinct jump separating the high densities in the molecular cloud and the low densities in the \hii\ region.  A dotted contour is drawn as reference to the jump in some figures including Fig.~\ref{fig:results}, left panel, enclosing densities of more than $10^{5}\qcm$ with an uncertainty better than 1\,dex.  This discontinuity may partly be the result of the modeling being limited to three input quantities and the model properties. To understand the appearance of the discontinuity, the general anatomy of the reduced $\chi^2$ in the $n$-{\GUV}-plane needs to be explained.

The colored solid lines in Fig.~\ref{fig:chi2} mark the {\nG}-pairs predicting the observed line ratios and \ifir, respectively, with colored dotted lines indicating the 1$\sigma$ uncertainty of each quantity. The observed FIR intensity (red) defines the range for the UV field intensity, but sets no constraint on the H-density (horizontal \ifir\ line). The {\nG}-pairs predicting the observed \CO\ to \oi146\um\ ratio (green) lie on an L-shaped line. A high ratio requires high densities and UV radiation fields, while low densities or low UV fields lead to a low \CO\ to \oi146\um\ ratio. The intersection of the green with the red line would set the density, if we  disregarded the \oi146\um\ to \ifir\ ratio (blue). The {\nG}-pairs satisfying a given \oi146\um\ to \ifir\ ratio form an oval. In the PDR, where we observe  higher \oi146\um\ to \ifir\ ratios, the oval is relatively small. Together with the high \ifir\, there are no intersections of the blue and red line, but their $1\sigma$ neighborhoods intersect. The minimum $\chi^2$ is found in this intersection region on the green line resulting in a H-density around $10^6\qcm$. Position~1 (peak of the continuum emission and most PDR transitions) and position~2 (a representative location in the PDR with relatively weak \CO\ emission) in Figs.~\ref{fig:chi2} and \ref{fig:results} are examples of how the high densities come about.

\edit1{Towards the \hii\ region, the \CO\ line gets weaker and is not detected at all away from the molecular cloud (except around position 6, which will be discussed in Sect.~\ref{sec:pillar}). At position 4 the \CO\ to \oi146\um\ ratio is so low that the lower dotted $1\sigma$ line for this ratio transition is missing in that panel in Fig.~\ref{fig:chi2}.  For most of the \hii\ region including position 5, the observed \CO\ to \oi146\um\ ratio is zero and the ratio's uncertainty becomes its upper limit. Only that $1\sigma$ upper limit can be plotted in the {\nG}-plane and the ratio becomes effectively an upper limit for the hydrogen density.}

The sharp drop from densities $10^6\qcm$ to about $10^4\qcm$ happens in the model because as \ifir\ and the ratio I(\oi146\um)/\ifir\ drop, there are two {\nG}-pairs that  match the measurements (the red line and the blue oval intersect twice), one at a density higher than $10^6\qcm$ and one around $10^4\qcm$. \edit1{Due to the \CO\ to \oi146\um\ ratio serving as an upper limit,} only the lower density solution is compatible with the observations.  This transition is illustrated by positions 3 and 4 in Figs.~\ref{fig:chi2} and \ref{fig:results}.

\edit1{While the PDRT derives a solution for the area where no \CO\ was detected, the applicability of the PDR model is questionable. The model assumes that we are looking at a PDR, but in the edge-on PDR in M17-SW, we are looking past the PDR in the upper parts of the maps, where no \CO\ but \oiii\ is detected. The sight lines cross the \hii\ region, but do not hit the PDR. Thus, the model may not even be  applicable there. The model-derived low densities should therefore seen only as upper limits for the densities in the \hii\ region. We postpone further discussion of the \hii\ region to a future paper.}

\subsection{Comparison to Literature}
\label{sec:comp-liter}

How do the derived H-densities compare to other density estimates?  \cite{Perez2015_densegas} derived densities for four locations in the molecular cloud. Their HCN and CO peak locations are close to our position~1. Their southern and western locations are within the area where the PDRT fitting leads to high density solutions. They estimate for the warm $H_2$ component, which is the dominant one, densities between $10^{5.7}$ and $10^6\qcm$ for these four locations.

We derived very similar densities in the molecular cloud, which we define here as the region where the modeling finds a high-density solution with an uncertainty better than 1\,dex excluding the area around the O-stars at the top of the map. The range of densities in the molecular cloud is $10^{5.6}$ to $10^{6.1}\qcm$ (excluding 5\% of the pixels as outliers), with a median of $10^{5.9}\qcm$ matching the findings of \cite{Perez2015_densegas}.  Our densities also fall into  the range of densities found by \cite{Stutzki90} for the clumps in their clumpy PDR model ($10^{4.5}$ to $10^7)\qcm$. Using pure-rotational lines of H$_2$ in the mid-infrared, \cite{Sheffer2013} even find densities of a few $10^7\qcm$ adding that the H$_2$ emission is produced ``in high-density clumps immersed in an interclump gas of density lower by two or three orders of magnitude''.

\cite{Meixner92} analyzed FIR fine-structure and molecular lines observed by the Kuiper Airborne Observatory (KAO) with lower spatial resolution than our maps obtained with SOFIA. In their paper, the density and the UV intensity is derived for four positions. Their first position is about 15\arcsec\ east of our position 1. Our position 2 is about 15\arcsec\ south of the midpoint between their second and third positions, which are separated by 40\arcsec\ in a nearly east-west direction. Their fourth position is at the south-eastern edge of our maps. All their position fall into the region, where our method finds a high density solution. The densities derived from the \cii\ to \oi146\um\ ratio only by \cite{Meixner92} are about a factor of four below our densities at these positions with a similar relative trend between the positions. Later in that paper, \cite{Meixner92} adopt, similar to \cite{Stutzki90}, a clumpy PDR model with a density of $10^{5.7}\rm\,cm^{-3}$ in the clumps embedded in a core with a density of $10^{3.5}\rm\,cm^{-3}$.

%% Dropped paragraph in response to the referee:
%% In the \hii region, which we define here as the region where the modeling finds a low-density solution with an uncertainty better than 1\,dex and where we detected \oiii\ emission, the median density is $10^{3.8}\qcm$. 

\subsubsection{Predictions for the \oi63\um\  and \cii\ lines}
\label{ssec:predictions}

Having estimates for the density and UV intensity, the PDRT can be used to predict line rations containing  the \oi63\um\ and \cii\ lines. This approach was inspired by the PDR analysis of FIFI-LS observations of the circumnuclear ring in the galactic center by \cite{Iserlohe2019}. As reference to these optically thick lines, we use the optically thin \oi146\um\ line.

The predicted and observed \cii\ to \oi146\um\ ratios are shown in the left panel of Fig.~\ref{fig:ci_oi_ratios}. Where the density is high, a fairly constant ratio of around 0.4 is predicted for \cii/\oi146\um. The observed ratio is also fairly constant, but around 1.5 rather than 0.4.
\edit1{Towards the \hii\ region, the observed line ratio rises. The predicted does, too, but the model may not be applicable there. Towards the molecular cloud, we observe about 3 times more \cii\ emission relative to \oi146\um\ than predicted by the PDRT}, even if we subtracted a foreground \cii\ emission. Following the example of \cite{Stutzki}, who subtracted a foreground \cii\ emission of 0.5\merg before comparing their PDR model to the observations, we could subtract also 0.5 or 0.9\merg, which is the lowest intensity in our \cii\ map. That would lower the observed \cii/\oi146\um-ratio from 1.5 to about~1. Still, the \cii\ emission from the PDR is a factor of~2 higher than the predicted value relative to \oi146\um.

A clumpy PDR model \citep[see, e.g., ][]{Stutzki90,Meixner92,Sheffer2013} would predict a stronger \cii\ emission from the PDR, but that cannot be the whole explanation. A clumpy PDR model would not explain the higher \cii/\oi146\um\ ratio in the \hii\ region. A combination of a widespread foreground \cii, under-predicted \cii\ due to a non-clumpy model, and \cii\ coming from other local phases not included in the simple model \citep[see also][]{Perez2015_atomicgas} may all contribute to the discrepancy to the observed \cii\ emission.

The \oi\ lines are much more confined to the PDR and the optical depth effect, which excluded the 63\um\ line from the PDR modeling, should become apparent when comparing the predicted and observed line ratios (right panel of Fig.~\ref{fig:ci_oi_ratios}). \edit1{The ratio is predicted to be roughly constant in the high density region at around~30.  In contrast, the observed ratio clearly drops from high though uncertain values of above 10 in the \hii\ region down to less than two and even to unity in some places deep into the cloud.} The increasing self-absorption of the \oi63\um\ line deeper into this edge-on PDR is exactly what we expected to see in this edge-on geometry. It also matches the observations by  \citet{Perez_Ringberg2017}, which show no sign of self-absorption in the high resolution \oi63\um\ spectra in the \hii\ region but strong self-absorption in the molecular cloud. 

\subsection{PDR Structure}
\label{sec:structure}

According to the literature and our analysis, the PDR is clumpy and allows the UV to penetrate the molecular cloud relatively deeply exciting more C$^+$ than in a homogeneous medium. Still the PDR looks edge-on as seen by the layering in Figs.~\ref{fig:fs_lines} and \ref{fig:structure}, but there is also some overlap of the  ionized and molecular phase as traced by, e.g., the ionized and atomic oxygen emission. There is also the big clump hosting M17-UC1 protruding very obviously into the \hii\ region, and that clumpiness and projection effect seen there would also be expected down to at smaller spatial scales. So, the overlapping could be explained alternatively by a projection effect as in the model by \cite{Sheffer2013} (also a clumpy PDR model) where the PDR surface is behind the molecular material from our vantage point to explain extinction effects.

Can we still identify the locations of the ionization and photo-dissociation fronts? While the density jump in the PDR-model indicates the magnitude of the density ratio between the \hii\ region and the molecular cloud, its abruptness may be an artifact of modeling with just the three input quantities and, thus, its exact location is dependent on the model parameters, too.

The left panel of Fig.~\ref{fig:structure} shows a three-color image of M17-SW in three mid-infrared tracers. There is a color change showing where the conditions change in this layered edge-on PDR. There is the red-blue region with strong 8\um\ and \oiii52\um\ emission with a quick transition to the green-blue region with strong 37\um\ and some 8\um\ but vanishing \oiii52\um\ emission. Similarly, the near infrared H$_2$ and Br$\gamma$ observations by \cite{Burton2002} show a sharp transition from ionized to molecular material. The H$_2$ and Br$\gamma$ both form ridges next to each other, especially prominent as layers on the clump containing M17-UC1 (lower right panel of Fig.~4 in \citealt{Burton2002}).

The yellow line in Fig.~\ref{fig:structure}, denoting the density jump, follows the above mentioned color change fairly well, and thus also the transition from the \hii\ region to the molecular cloud, except around the clump containing M17-UC1. There, the yellow contour is well in front of the sharp edge of the clump towards the \hii\ region, which can be seen in the IRAC Band~4 and the FORCAST~37\um\ images in Fig.~\ref{fig:structure}. Since the PDRT modeling is based on the FIR maps smoothed to a resolution of about~17\arcsec\, the location of the density jump can easily be pushed out by a good fraction of the maps' resolution around the flux peaks of these maps, which all peak around M17-UC1. The pillar in the background (see Sect.~\ref{sec:pillar}) contributes further to an additional notch  in the contour into the \hii\ region.

Below the yellow line, the modeling tells us that the gas, even if clumpy, has mostly a density around $10^6\qcm$. The atomic layer of PDRs forming on the surface of molecular clouds has a hydrogen column density typically of about $2-4\times10^{21}\mathrm{cm}^{-2}$ \citep{Hollenbach99}. That means the atomic layer here has a thickness on the order of only~$10^{-3}$\,pc.
Further, \cite{Hollenbach99} report in their section on ``Time-dependent and nonstationary PDRs'' that for relatively high gas densities ($\approx10^6\qcm$) and UV intensities ($10^{4.5}G_0$), as derived here for the PDR region, the ionization and dissociation fronts can even merge as H$_2$ flows fast enough towards the \hii\ region. Thus in M17-SW, the ionization and photo-dissociation fronts on the clumpy PDR-medium may even be merged. Even with a clumpy structure and projection effects, the density jump in the model and the layered appearance of the various tracers indicate that both fronts should be where the red and yellow contours drawn in Fig.~\ref{fig:structure} are close together. On the M17-UC1 clump, the contours diverge as the geometry departs significantly from the a simple edge-on geometry due to the high density clump protruding into the \hii\ region.

\subsubsection{The pillar}
\label{sec:pillar}

A linear feature, best seen here in the IRAC4 data (Fig.~\ref{fig:structure}), extends east into the \hii\ region towards the most massive stars in M17, CEN1a and CEN1b. This feature, dubbed ``pillar'' by \cite{Lim2020}, can also be seen in the FORCAST images (Fig.~\ref{fig:m17-sw}). Being bright at 20\um\ and fading into the background from 37 to 70\um, \cite{Lim2020} argue that the pillar is ``an edge-on view of the interface between a ridge of dust and the ionizing and heating stars interior to [the \hii\ region]''.

The pillar is also detected with FIFI-LS. There is faint \oi146\um\ and strong \oi63\um\ as well as \cii\ emission (Figs.~\ref{fig:fs_lines} and \ref{fig:lines}) tracing the pillar. The PDR lines trace it further west than apparent in the IRAC4 or FORCAST images. It seems to extend out of the clump protruding into the \hii\ region, but that maybe a projection effect.

The PDRT finds high-density solutions along this pillar. Pos.~6 is a point on the pillar near the edge of the mapped area. The \ifir\ at Pos.~6 is relatively low and \oi146\um/\ifir\ ratio is quite high leading to a wide range of densities nearly satisfying these two observed quantities (see Pos.~6 in Fig.~\ref{fig:chi2}). Together with the relatively high \CO/\oi146\um\ ratio as CO is detected at Pos.~6 (but only with an SNR of~2), the model has to settle on a hydrogen density of around $10^6\qcm$. Most of the high-density extension is shown only in faint orange in Fig.~\ref{fig:results}, because the uncertainty of the derived density is nearly~1.25\,dex.

From the mid-infrared images, we know that the pillar is narrower than the spatial resolution of our PDR model parameter maps. Therefore, it is not too surprising that the detection of the pillar in the PDR model is rather weak. Still, the detection of PDR emission from the pillar and high densities in the PDR model along the pillar support the presence of a high-density ridge of gas and dust irradiated by the nearby O-stars from the outside.

\section{Conclusions}

The PDR known as M17-SW has been mapped with SOFIA/FIFI-LS in several fine-structure lines and high-J CO lines. The line ratio maps of the optically thin PDR-lines and the continuum maps, in form of the infrared intensity, were used to model the conditions across the mapped area using the PDRT tool, yielding a hydrogen nuclei density and UV radiation map. The UV map follows closely the FIR intensity map as the model assumes that all of the UV radiation is absorbed and re-emitted in the infrared. \edit1{The average H-density derived for the molecular cloud is $10^{5.9}\qcm$. The high H-densities and derived UV field point to the ionization and photo-dissociation fronts being nearly merged. While a density jump is expected from the \hii\ region into the PDR, the sharp jump between these two density regions may be partially an artifact of the modelling with only three input quantities \edit1{and the PDR model being not applicable in the \hii\ region.} A clumpy PDR-model would likely predict a brighter \cii\ emission than predicted by PDRT}, but it would not fully explain all the \cii\ emission as we see a ubiquitous \cii\ foreground not associated with the PDR.

However, using a relatively simple PDR model with optically thin tracers  allowed us to derive maps for the parameters of the model and to make predictions for the optically thick lines producing a consistent picture of the physical conditions in the PDR. The exact location of the jump in the derived H-density map may not be showing the location of the ionization and photo-dissociation front, but it approximates the area where the atomic layer in the edge-on PDR should be expected. The same area is indicated by the layered and partly overlapping emissions from the ionized, neutral, and molecular species. Thus, we have localized the ionization and photo-dissociation fronts in M17-SW.

A more detailed PDR-model also taking into account spectrally resolved observations of the optically thick lines should be able to better model the clumpiness and dynamics of the M17-SW PDR. It will be a challenge to model not only single points, but to create spatial maps of the model parameters as we have done here with the simple model, but it will bring us closer to a deeper understanding of the feedback mechanisms which regulate the star formation process.

\begin{acknowledgements}
Based on observations made with the NASA/DLR Stratospheric Observatory for Infrared Astronomy (SOFIA). SOFIA is jointly operated by the Universities Space Research Association, Inc. (USRA), under NASA contract NAS2-97001, and the Deutsches SOFIA Institut (DSI) under DLR contract 50 OK 0901 to the University of Stuttgart. Financial support for this work was provided by NASA through USRA (NNA17BF53C subcontract 04-0049). We thank A. Karska (Nicolaus Copernicus University, Poland) and an anonymous reviewer for constructive comments. This research has made use of  NASA’s Astrophysics Data System and the SIMBAD database, operated at CDS, Strasbourg, France. 
\end{acknowledgements}

\vspace{5mm}
\facilities{SOFIA}

\bibliographystyle{aasjournal}
\bibliography{m17,sofia,astro,pdr}

\appendix
\section{Maps}
\label{sec:maps}
In this appendix, all the maps listed in Tab.~\ref{tab:values} are displayed separately. For all maps but the \CO\ map, the orignal map is displayed on the left. On the right is the same map smoothed to the beam size of the map with the lowest spatial resolution, which is the \CO\ map. The \CO\ map does not need to be smoothed further. To the right of it is the FIR intensity map derived from the observed continua and Herschel maps.

The color bar next to each figure indicates the measured intensities for each line and the total FIR continuum. In the maps the color saturation fades for a SNR lower than 5 and reaches white at an SNR of 1 and lower. Contours are drawn at $\frac{1}{6}, \frac{1}{3}, \frac{1}{2}, \frac{2}{3}$, and $\frac{5}{6}$ of the peak flux, which is listed in \ref{tab:values}.

The stars and crosses mark the locations of the O-stars identified by \cite{Hoffmeister2008}. Stars mark spectral types earlier than O9 and crosses mark types O9 and O9.5. The diamonds indicate the positions 1 through 6 discussed in the paper. The circle in the lower left corner indicates the beam size. The arrow in the upper right corner points north. The reference position is the location of hypercompact \hii\ region M17-UC1 \citep{Sewilo2004}.

\begin{figure}
  \includegraphics[width=.5\linewidth,viewport=61 11 477 323]%
  {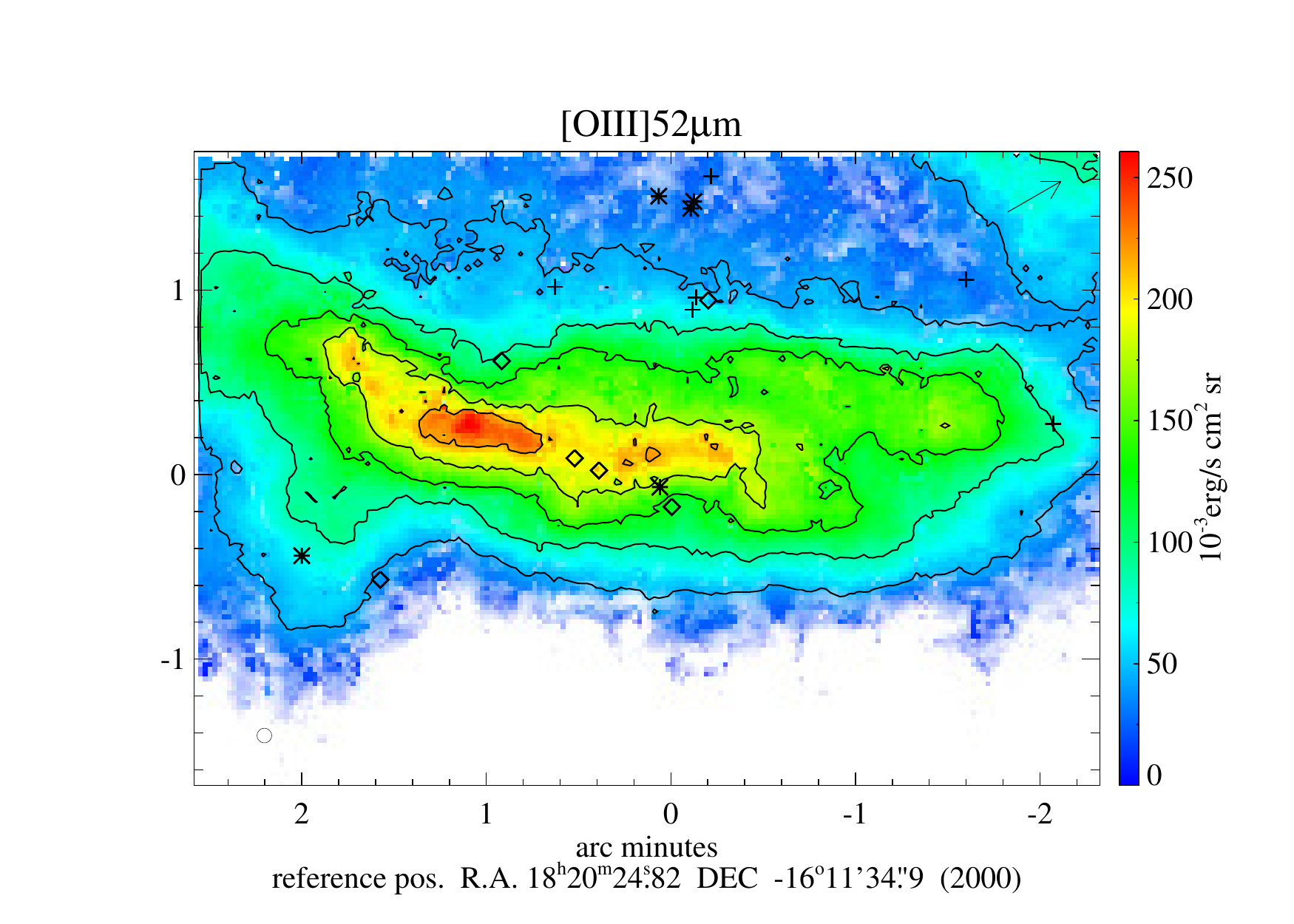}
  \includegraphics[width=.5\linewidth,viewport=61 11 477 323]%
  {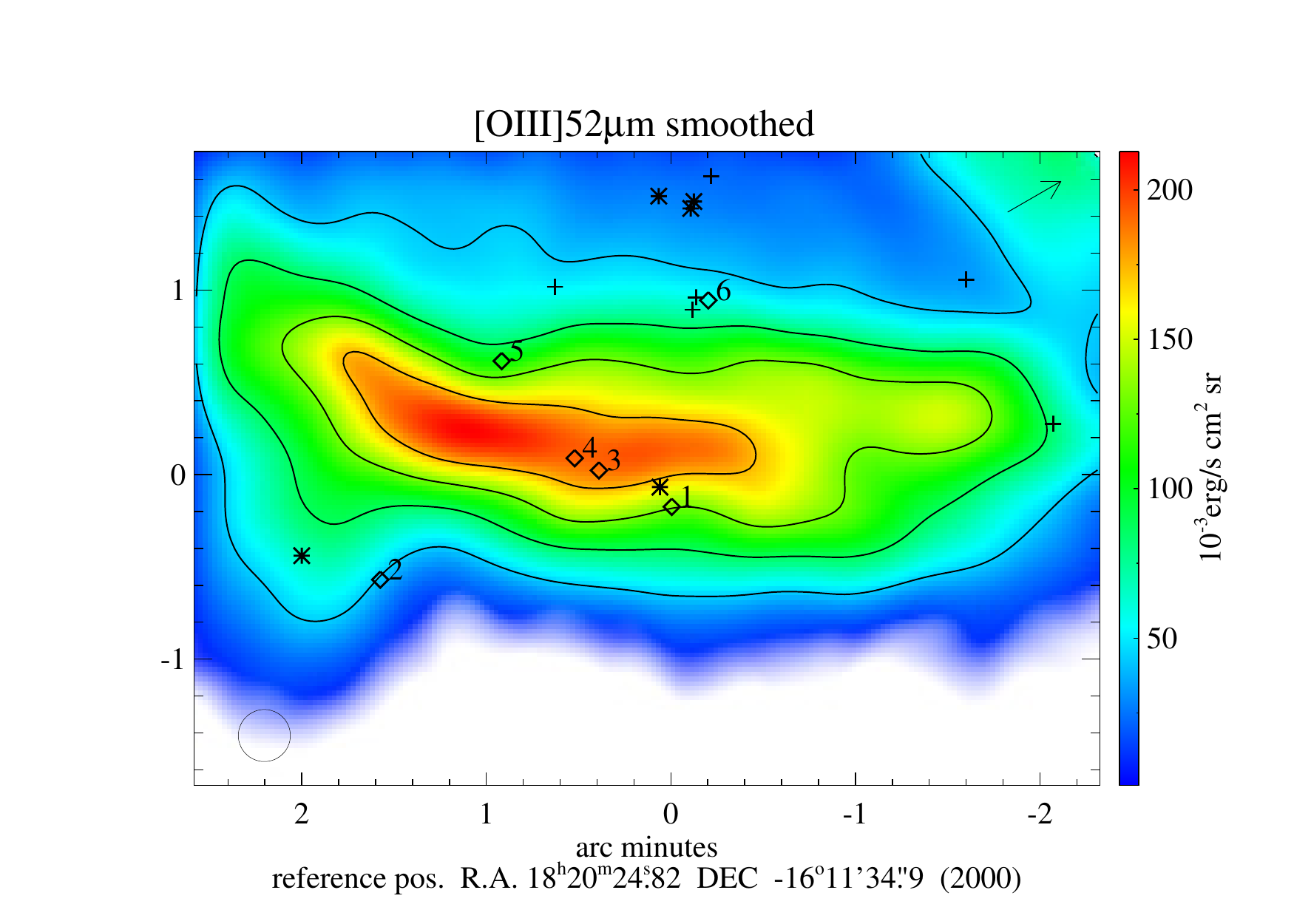}
  
  \caption{ \oiii52\um\ maps; see appendix~\ref{sec:maps} for details}
  \label{fig:oiii52}
\end{figure}

\begin{figure}
  \includegraphics[width=.5\linewidth,viewport=61 11 473 323]%
  {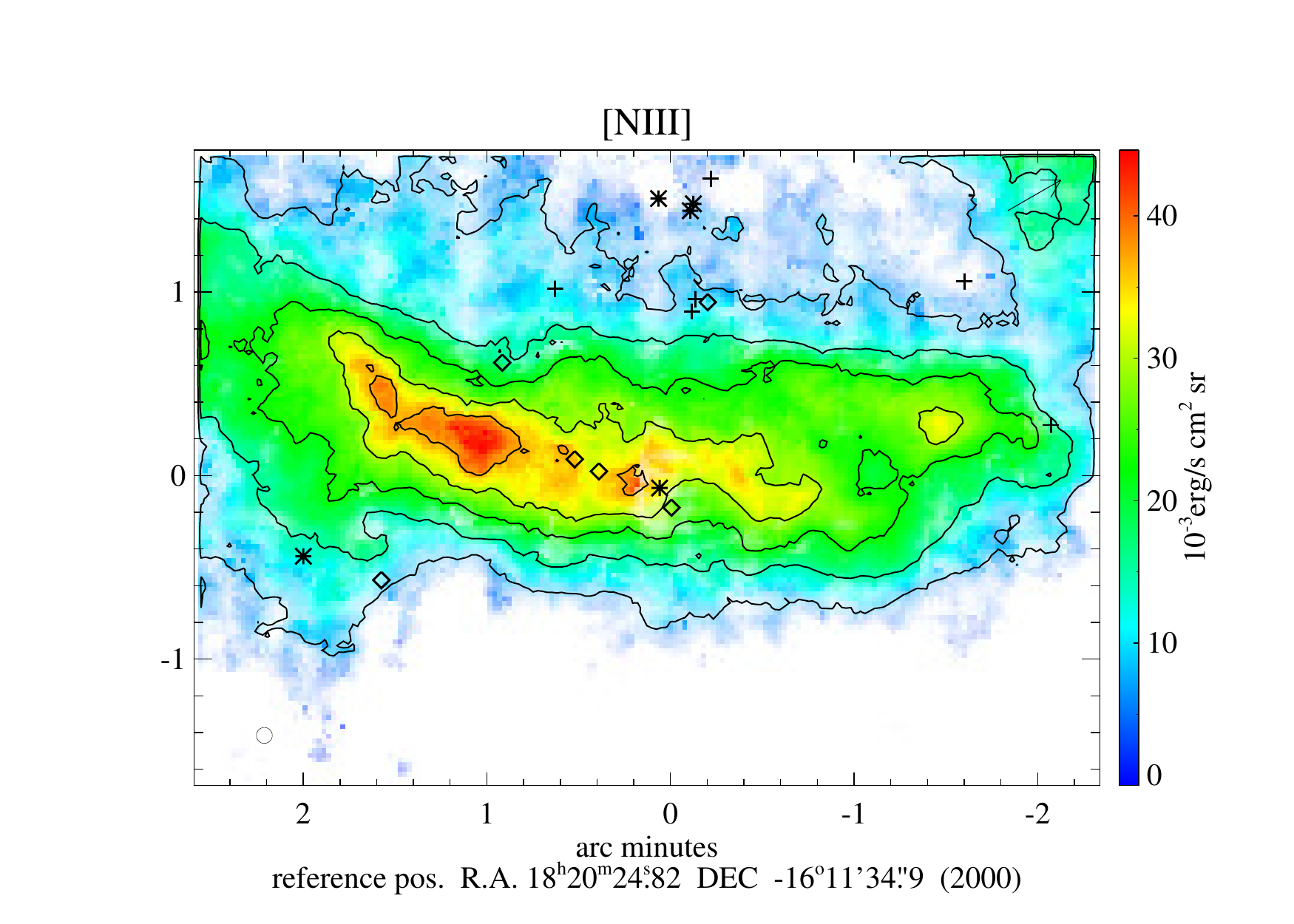}
  \includegraphics[width=.5\linewidth,viewport=61 11 473 323]%
  {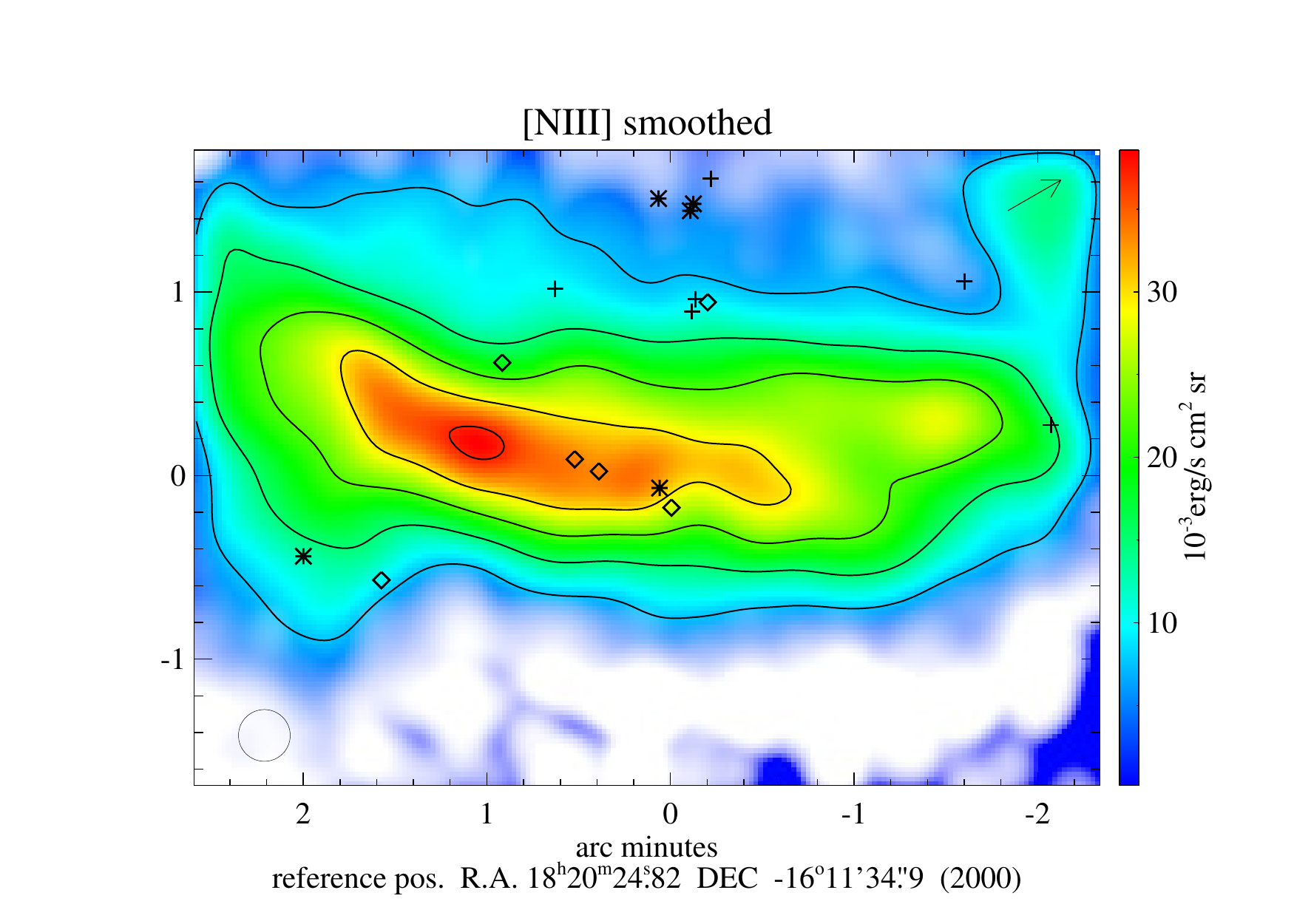}
  
  \caption{ \niii\ maps; see appendix~\ref{sec:maps} for details}
  \label{fig:niii}
\end{figure}

\begin{figure}
  \includegraphics[width=.5\linewidth,viewport=61 11 473 323]%
  {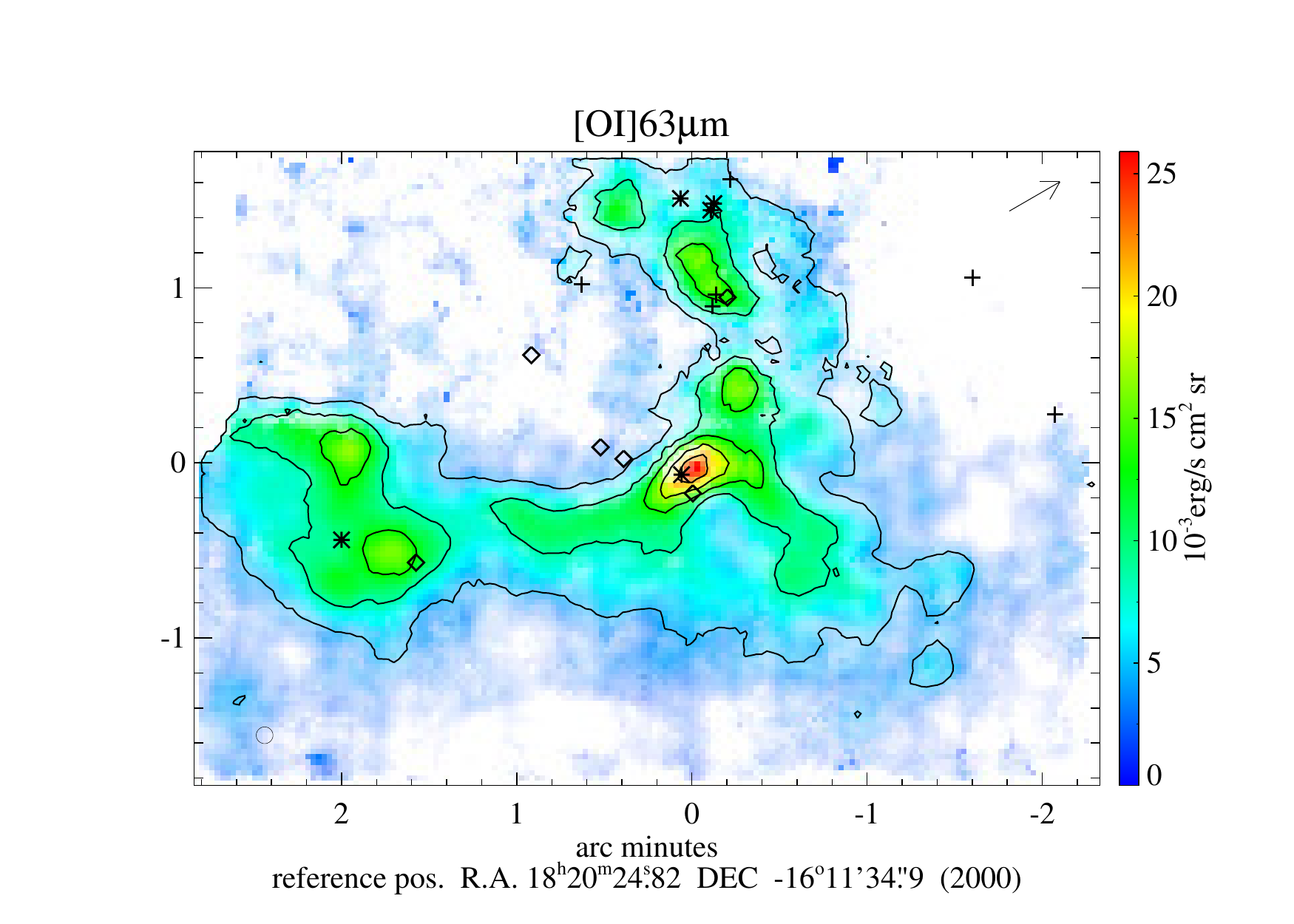}
  \includegraphics[width=.5\linewidth,viewport=61 11 473 323]%
  {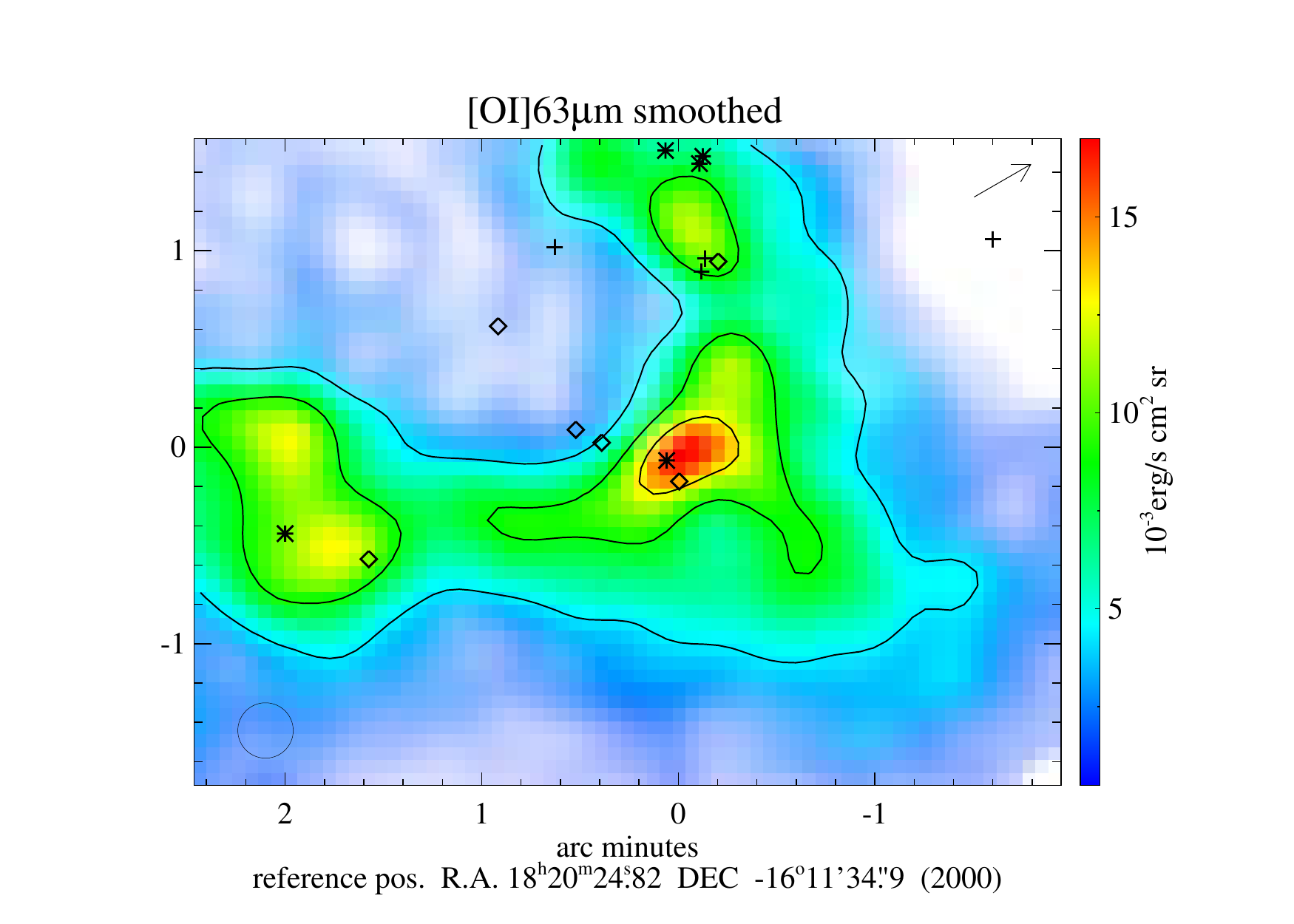}
  
  \caption{ \oi63\um\ maps; see appendix~\ref{sec:maps} for details}
  \label{fig:oi63}
\end{figure}

\begin{figure}
  \includegraphics[width=.5\linewidth,viewport=61 11 473 323]%
  {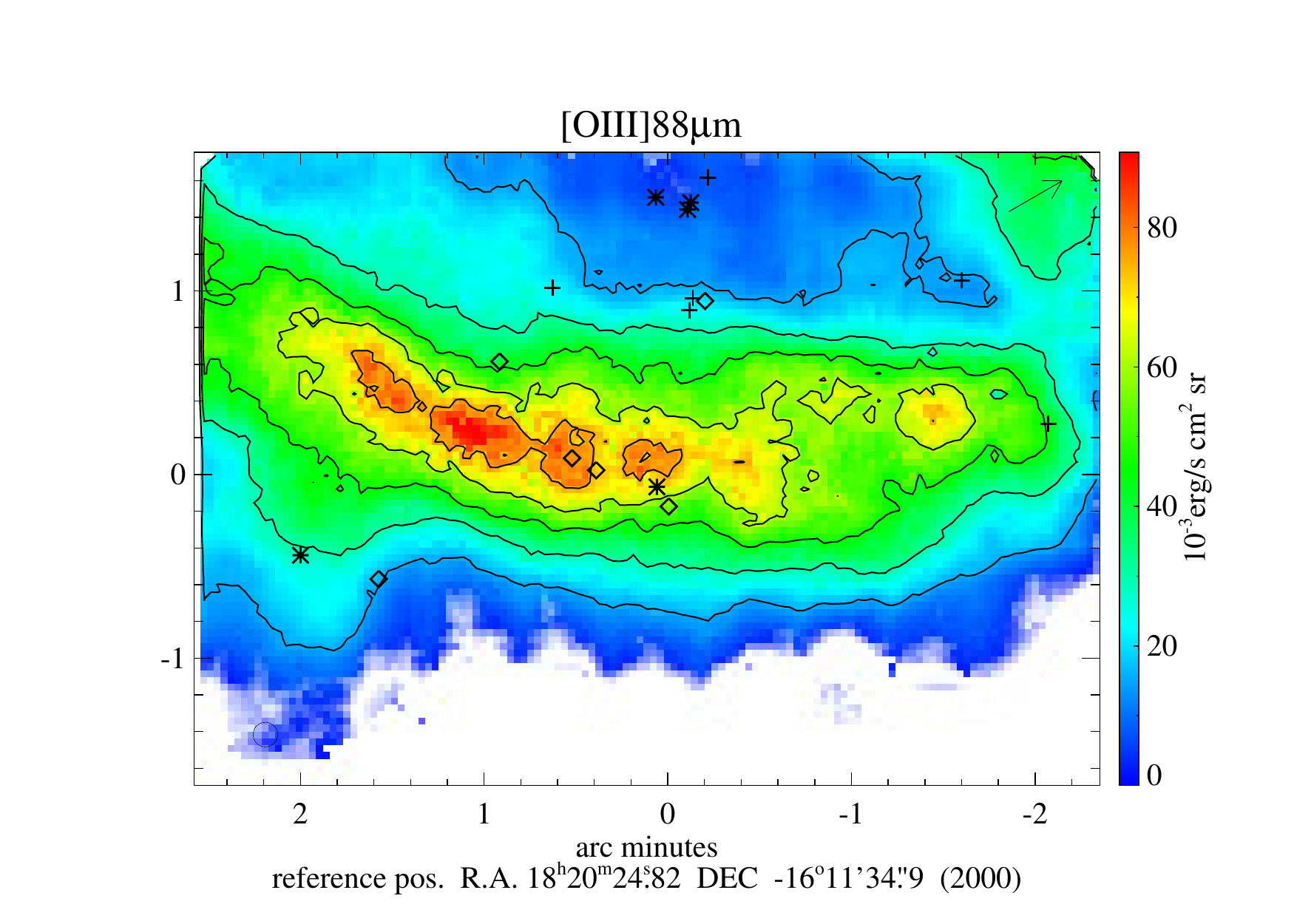}
  \includegraphics[width=.5\linewidth,viewport=61 11 473 323]%
  {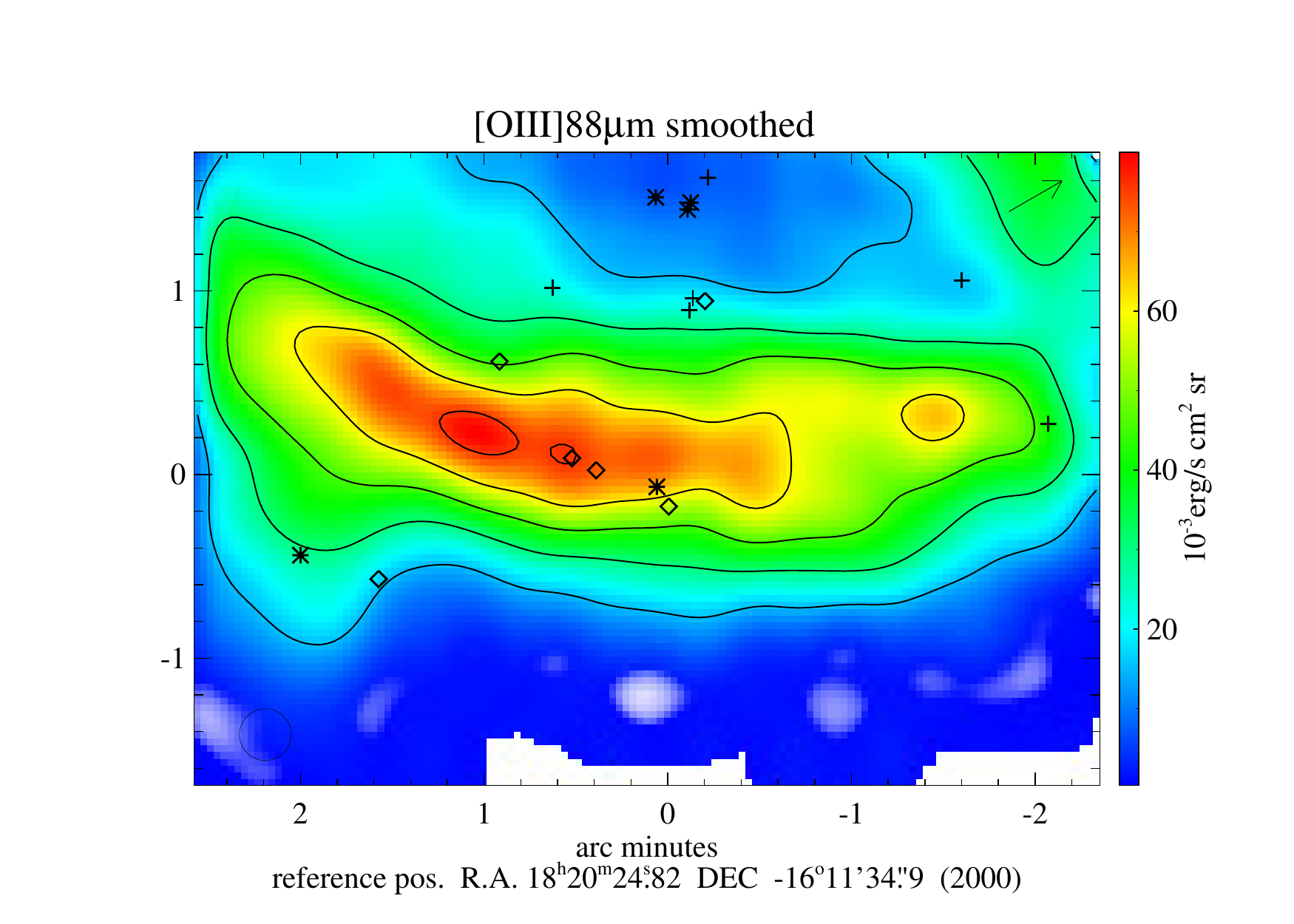}
  
  \caption{ \oiii88\um\ maps; see appendix~\ref{sec:maps} for details}
  \label{fig:oiii88}
\end{figure}

\begin{figure}
  \includegraphics[width=.5\linewidth,viewport=61 11 458 323]%
  {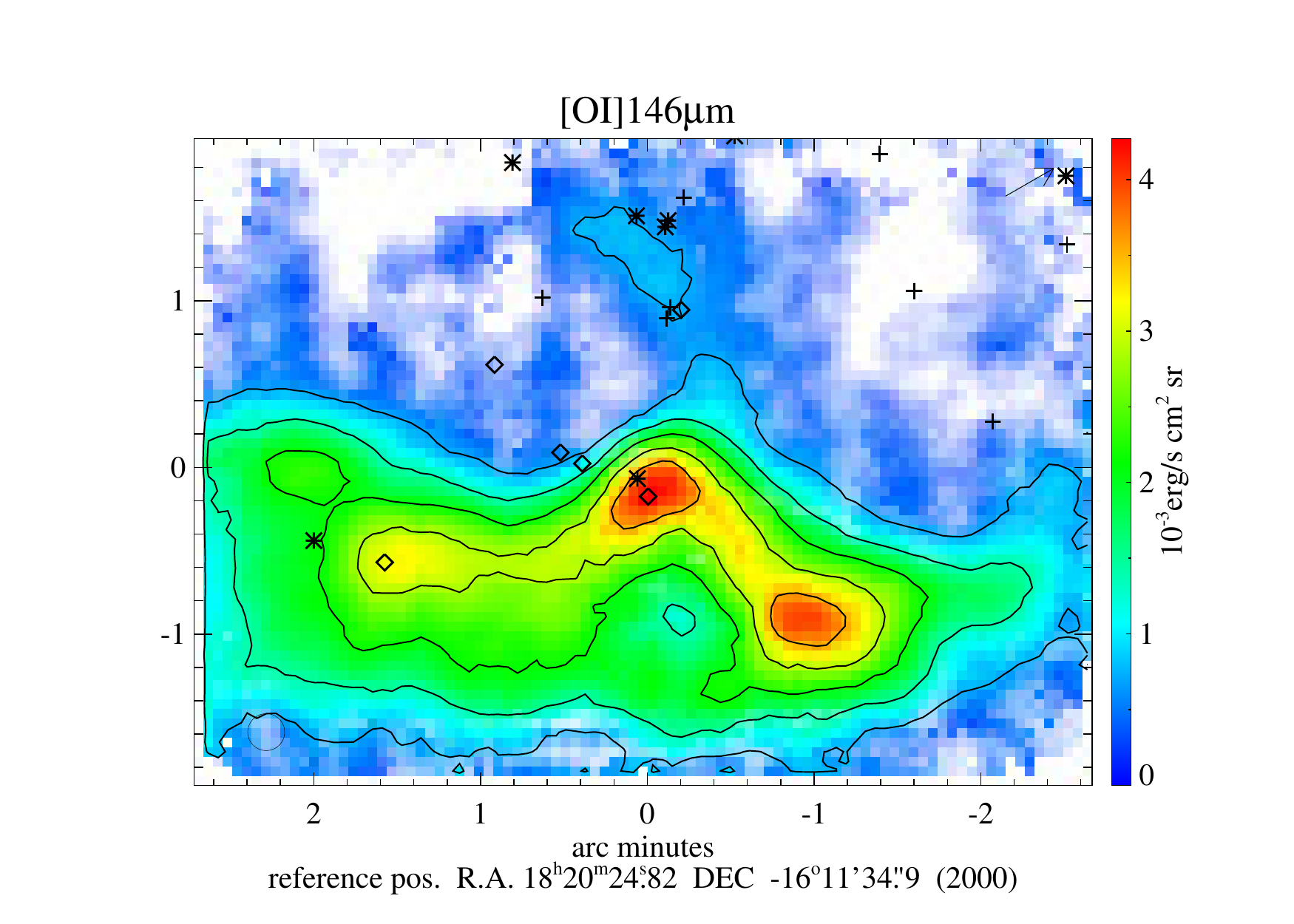}
  \includegraphics[width=.5\linewidth,viewport=61 11 458 323]%
  {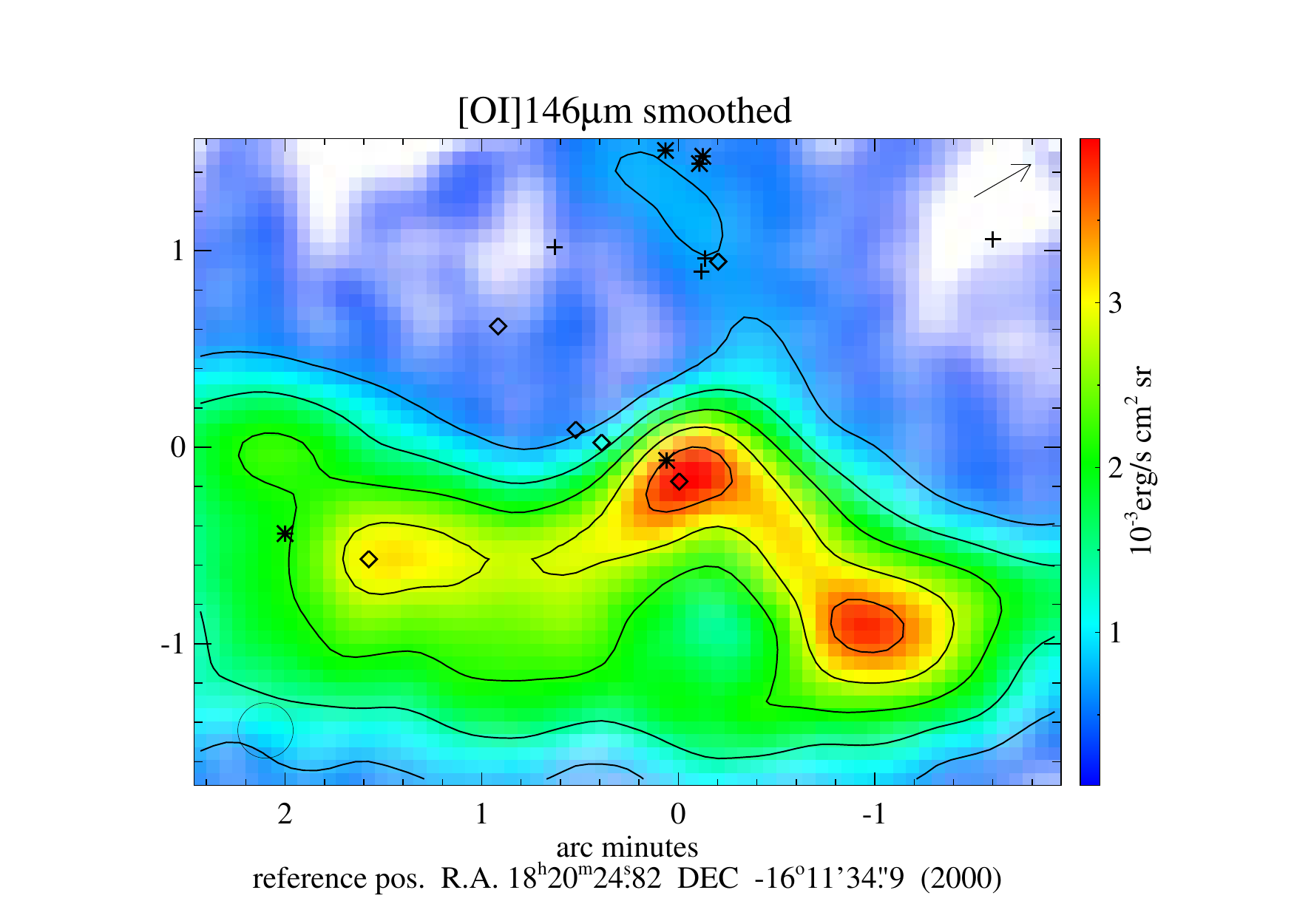}
  
  \caption{ \oi146\um\ maps; see appendix~\ref{sec:maps} for details}
  \label{fig:oi146}
\end{figure}

\begin{figure}
  \includegraphics[width=.5\linewidth,viewport=61 11 478 323]%
  {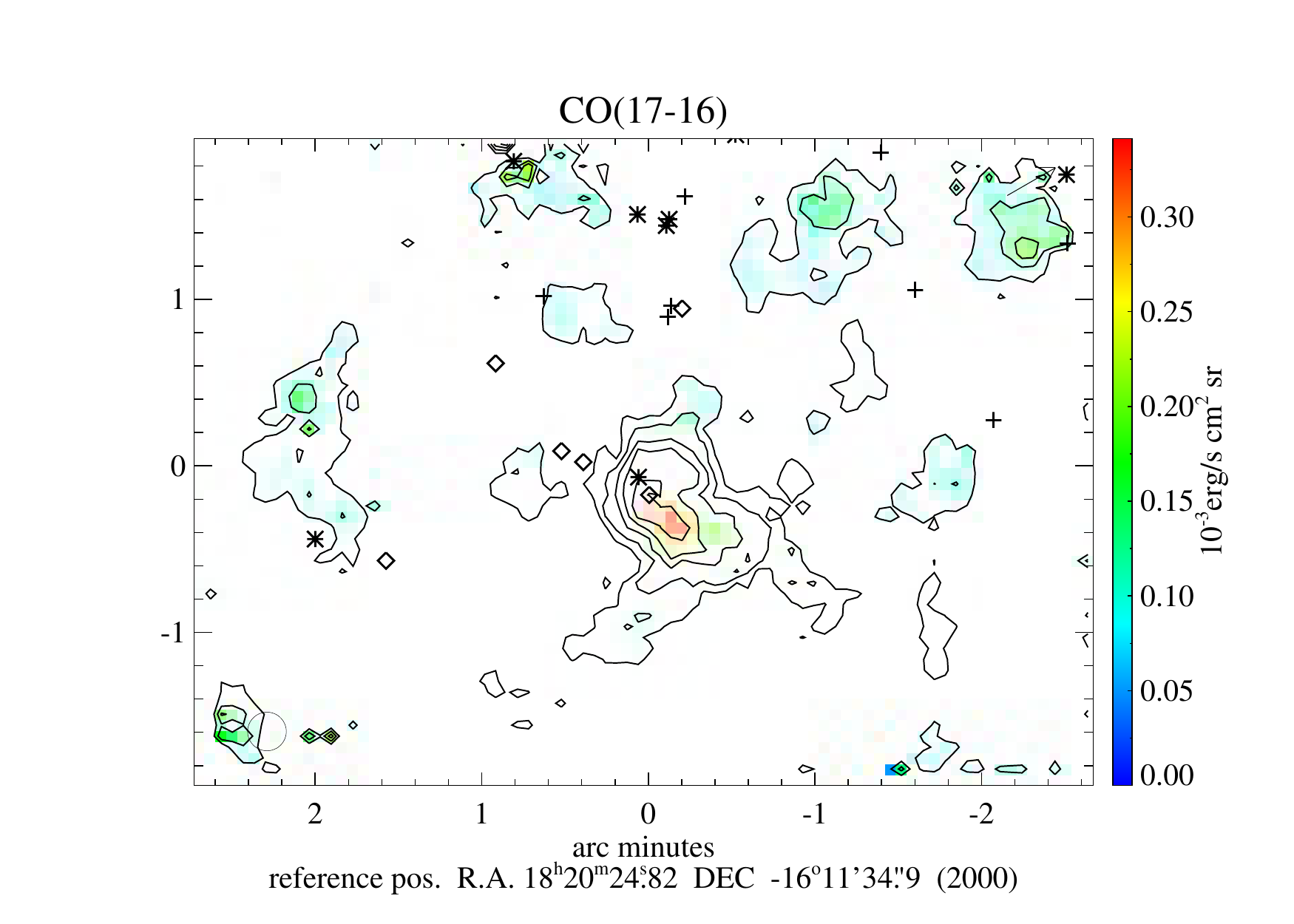}
  \includegraphics[width=.5\linewidth,viewport=61 11 478 323]%
  {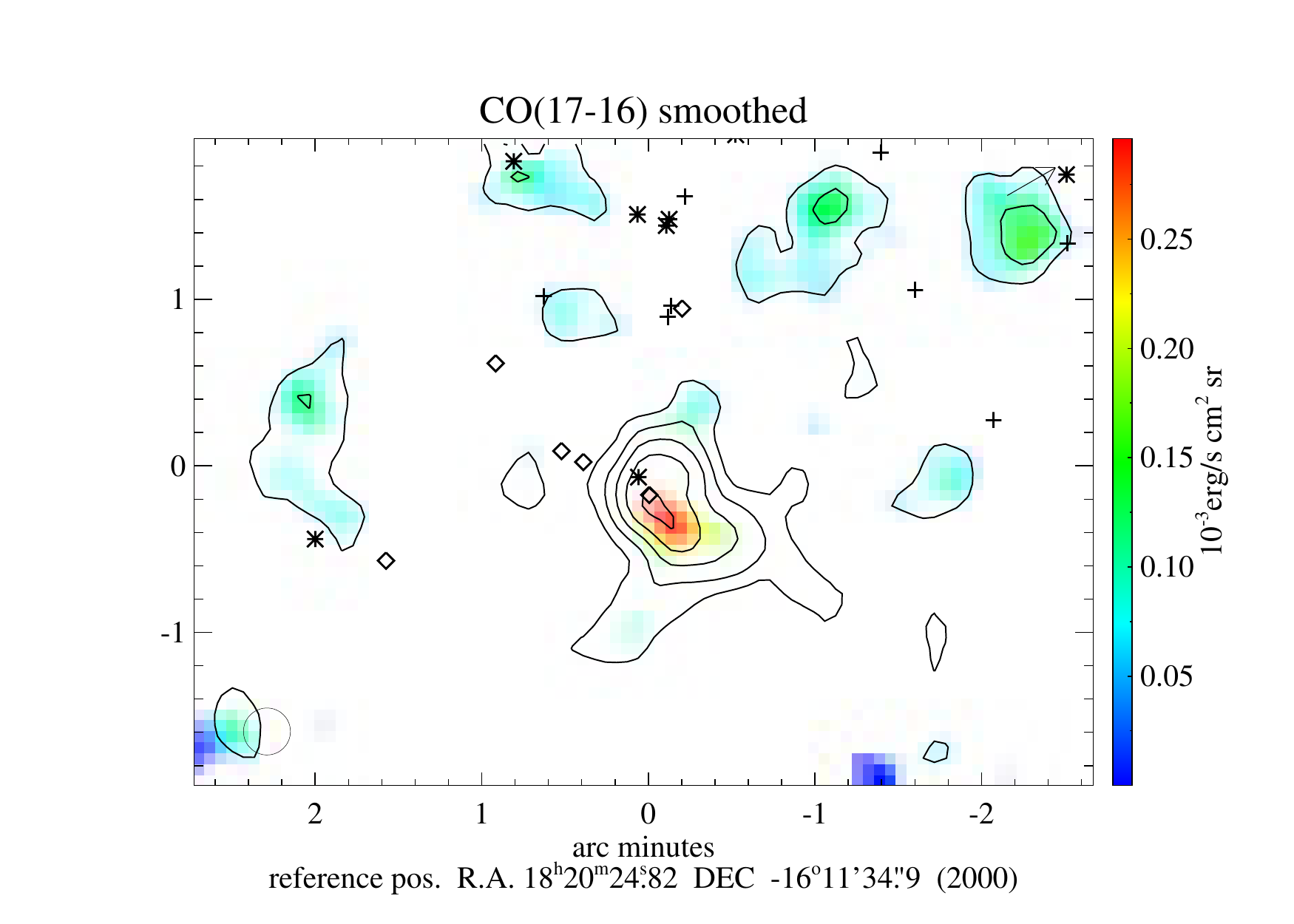}
  
  \caption{ CO(17-16)\ maps; see appendix~\ref{sec:maps} for details}
  \label{fig:co1716}
\end{figure}

\begin{figure}
  \includegraphics[width=.5\linewidth,viewport=61 11 458 323]%
  {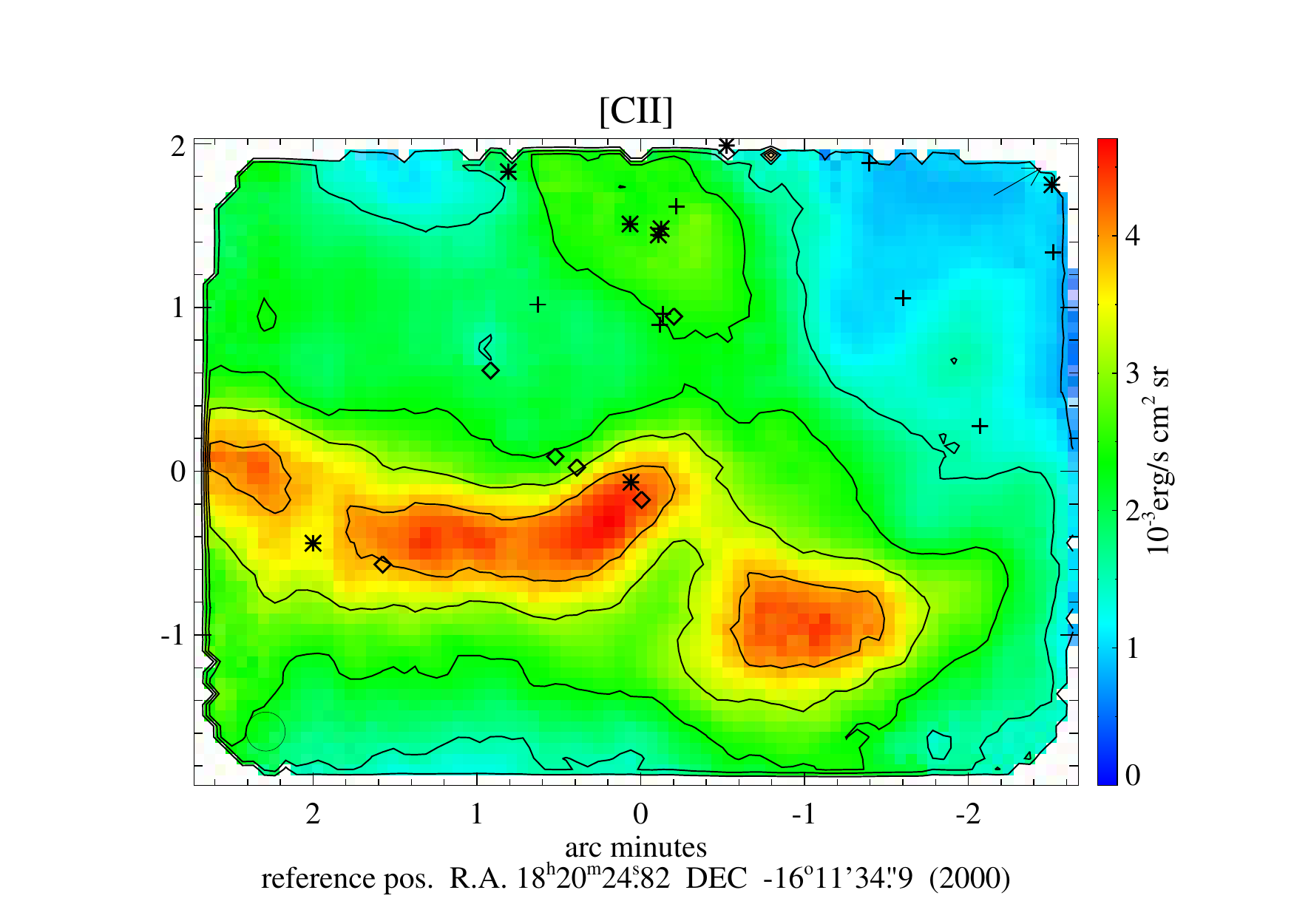}
  \includegraphics[width=.5\linewidth,viewport=61 11 458 323]%
  {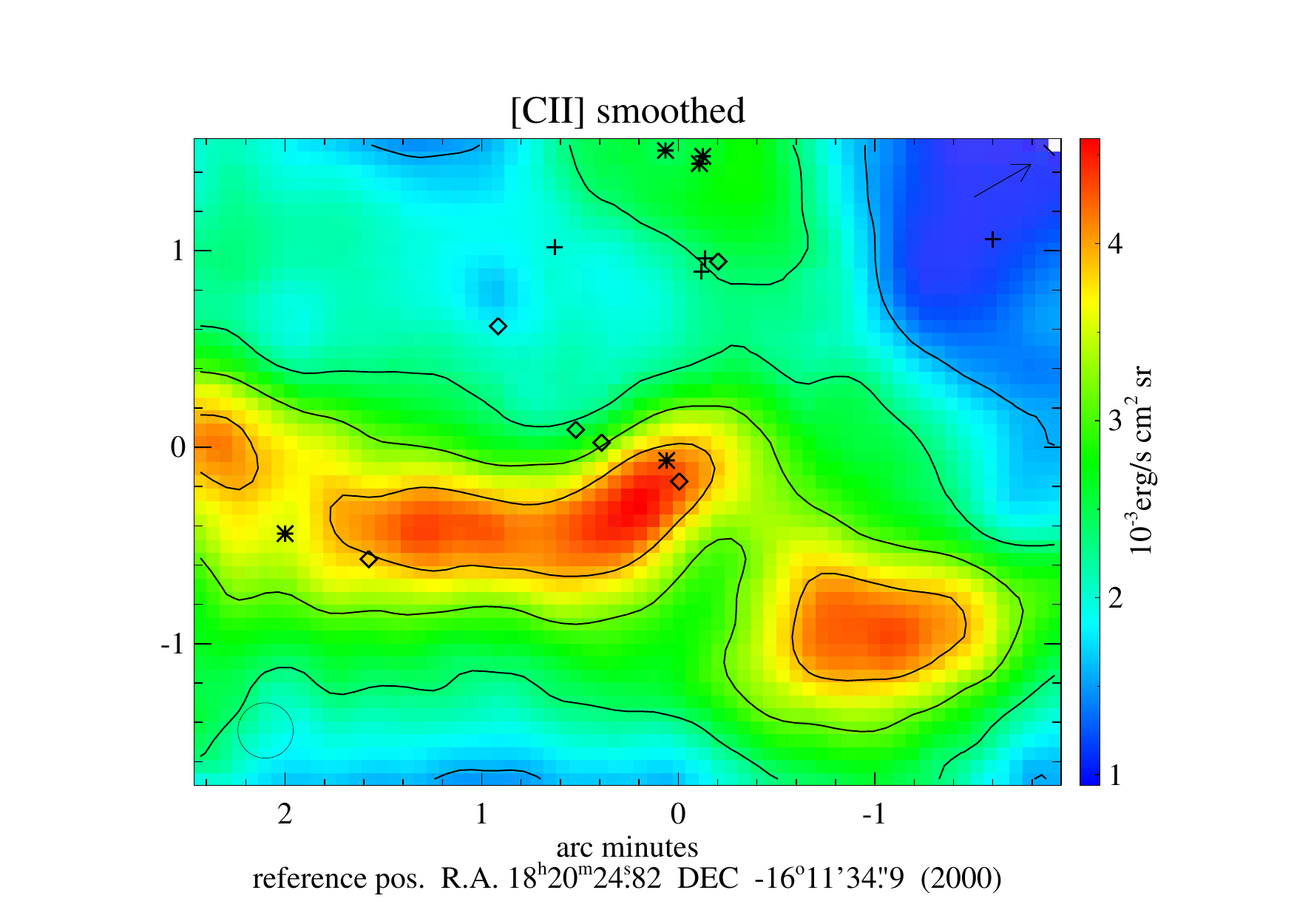}
  
  \caption{ \cii\ maps; see appendix~\ref{sec:maps} for details}
  \label{fig:cii}
\end{figure}

\begin{figure}
  \includegraphics[width=.5\linewidth,viewport=61 11 475 260]%
  {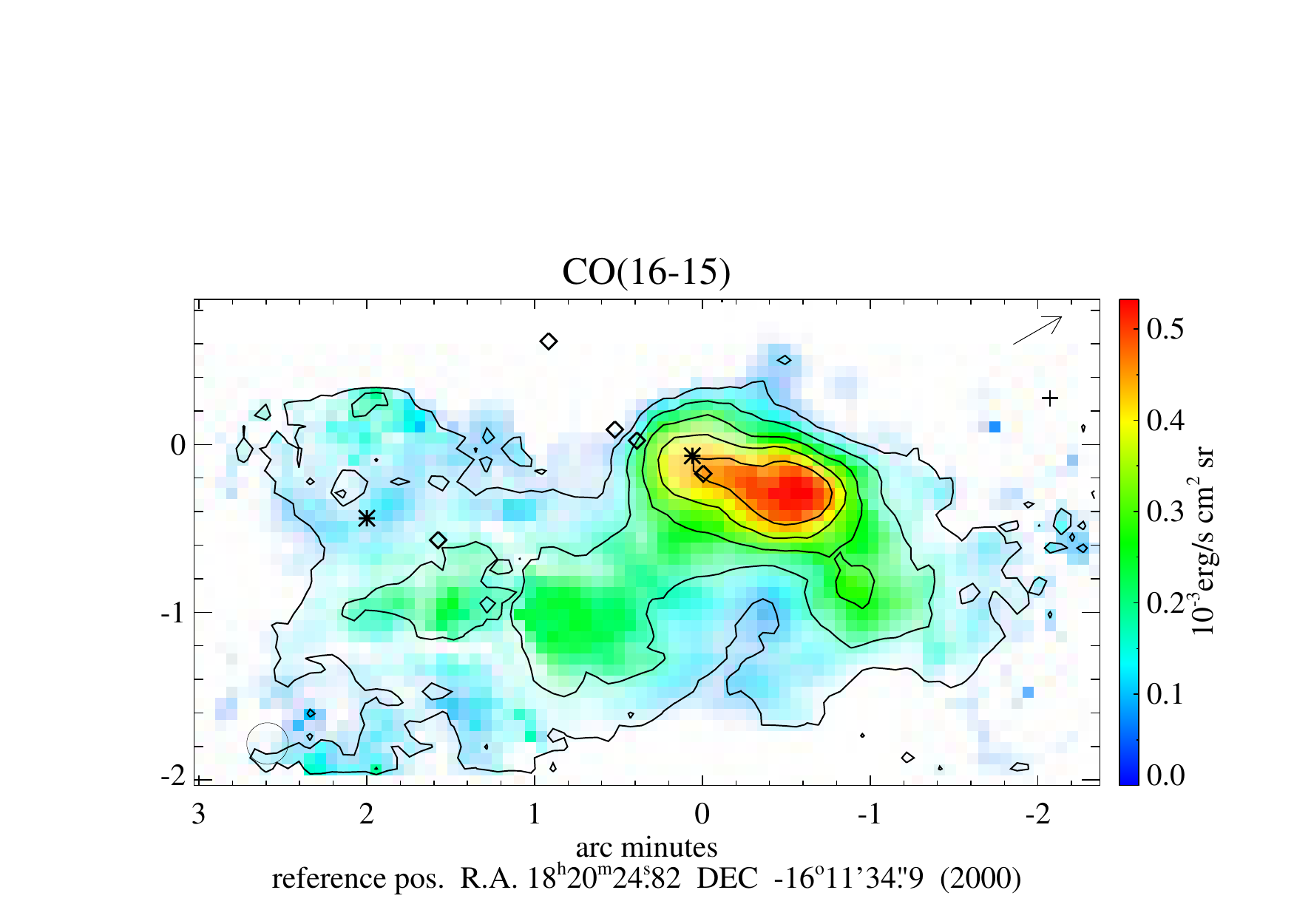}
  \includegraphics[width=.5\linewidth,viewport=61 11 475 260]%
  {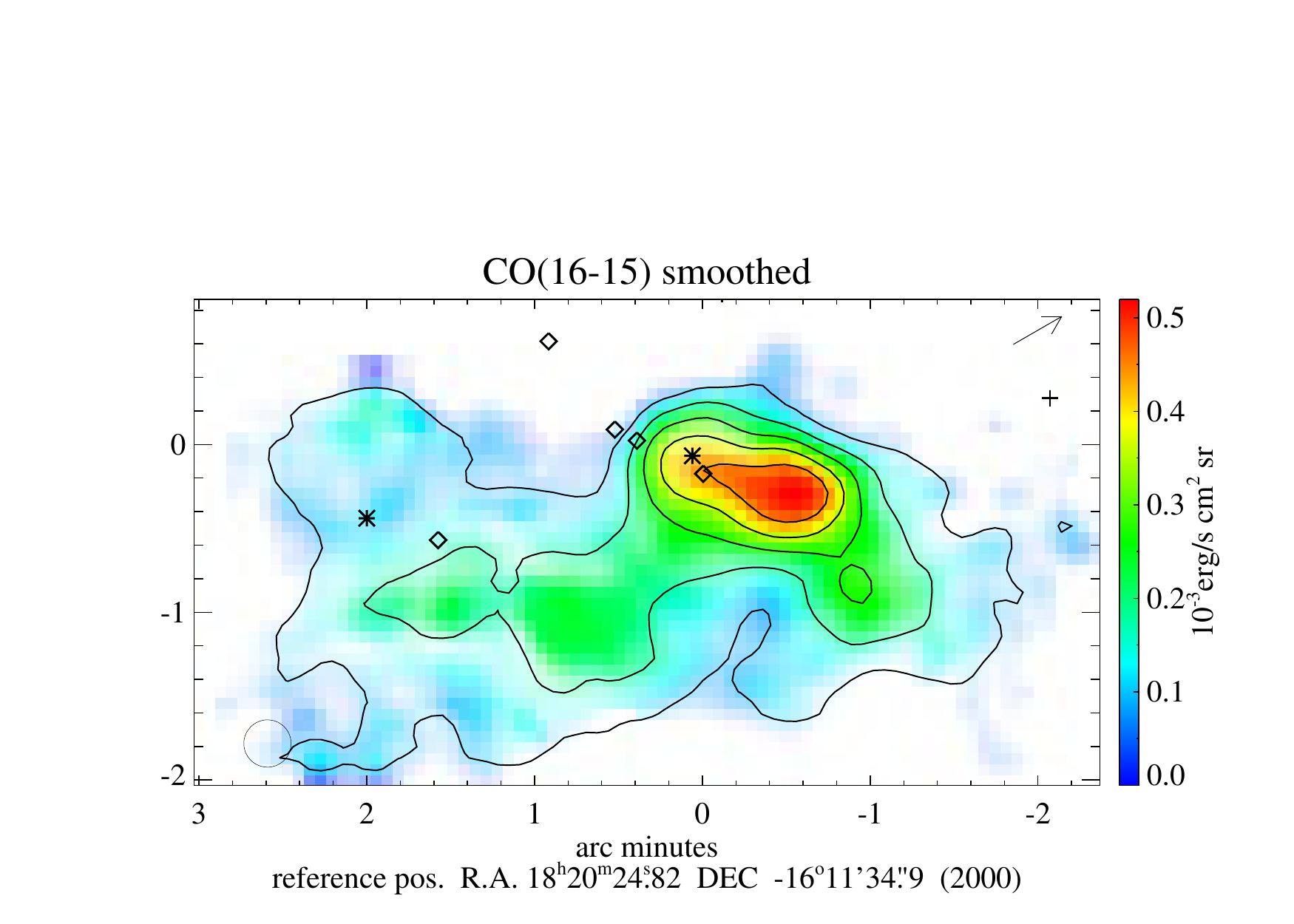}
  
  \caption{ CO(16-15) maps; see appendix~\ref{sec:maps} for details}
  \label{fig:co1615}
\end{figure}

\begin{figure}
  \includegraphics[width=.5\linewidth,viewport=61 11 460 323]%
  {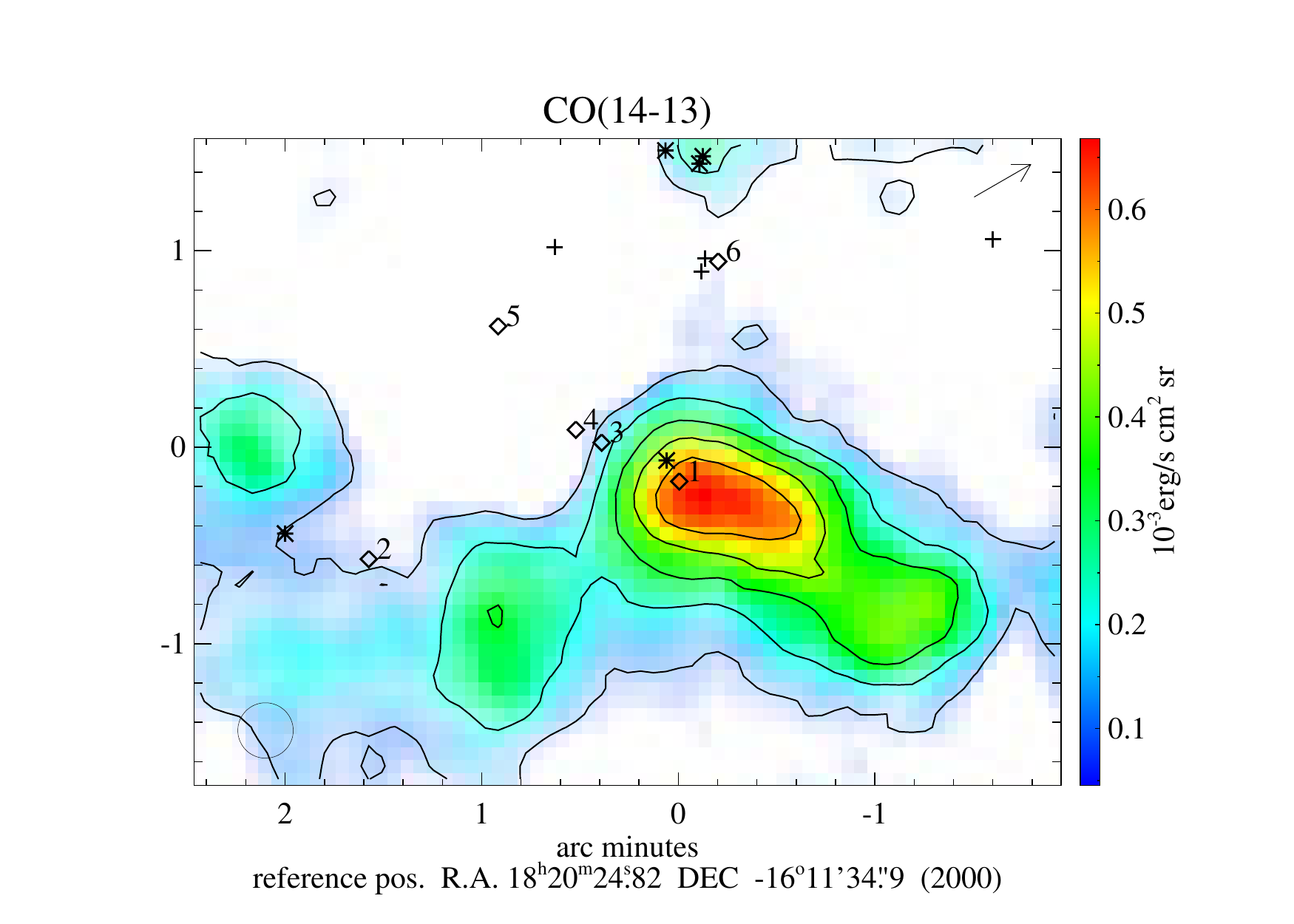}
  \includegraphics[width=.5\linewidth,viewport=61 11 482 323]%
  {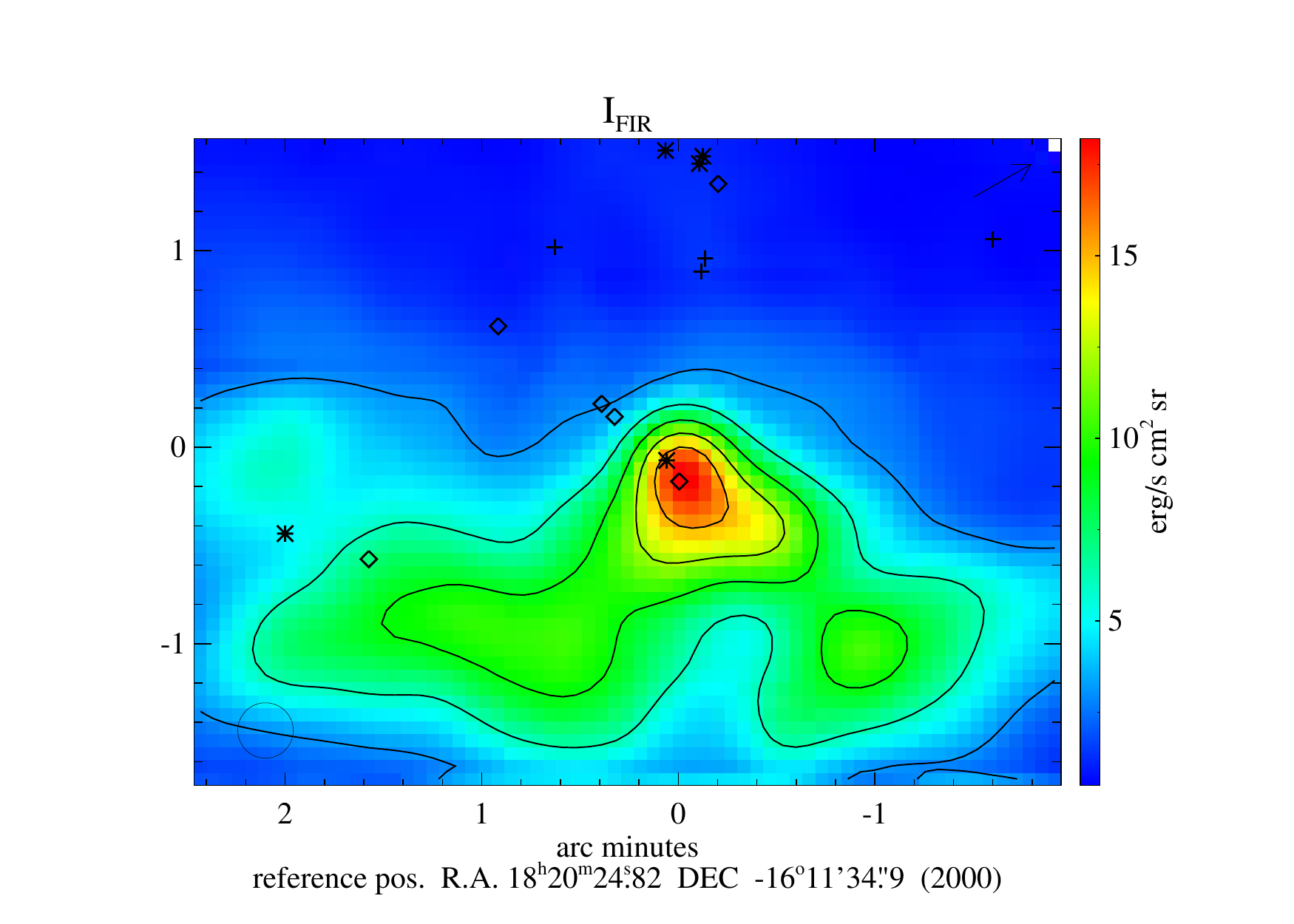}
  \caption{\CO and infrared intensity map; see appendix~\ref{sec:maps} for details}
  \label{fig:co_I_FIR}
\end{figure}
\end{document}